\def\year{2020}\relax
\documentclass[letterpaper]{article} 
\usepackage{aaai20}  
\usepackage{times}  
\usepackage{helvet} 
\usepackage{courier}  
\usepackage[hyphens]{url}  
\usepackage{graphicx} 
\usepackage{graphicx}  
\usepackage{amsmath}

\title{DeepBrain: Towards Personalized EEG Interaction through \\ Attentional and Embedded LSTM Learning}
\author{Di Wu$^\dag$, Huayan Wan$^\dag$, Siping Liu$^\dag$, Weiren Yu$^\ddag$, Zhanpeng Jin$^\S$, Dakuo Wang$^\P$
\\$^\dag$Hunan University, China 
$^\ddag$University of Warwick, UK 
$^\S$University at Buffalo, USA
$^\P$IBM Research AI, USA}

\begin{document}

\maketitle

\begin{abstract}
The ``mind-controlling'' capability has always been in mankind's fantasy. With the recent advancements of electroencephalograph (EEG) techniques, brain-computer interface (BCI) researchers have explored various solutions to allow individuals to perform various tasks using their minds. However, the commercial off-the-shelf devices to run accurate EGG signal collection
are usually expensive and the comparably cheaper devices can only present coarse results, which prevents the practical
application of these devices in domestic services. To tackle this challenge, we propose and develop an end-to-end solution
that enables fine brain-robot interaction (BRI) through embedded learning of coarse EEG signals from the low-cost devices,
namely DeepBrain, so that people having difficulty to move, such as the elderly, can mildly command and control a robot
to perform some basic household tasks. Our contributions are two folds:
1) We present a stacked long short term memory
(Stacked LSTM) structure with specific pre-processing techniques to handle the time-dependency of EEG signals and their
classification. 2) We propose personalized design to capture multiple features and achieve accurate recognition of individual
EEG signals by enhancing the signal interpretation of Stacked LSTM with attention mechanism. Our real-world experiments demonstrate that the
proposed end-to-end solution with low cost can achieve satisfactory run-time speed, accuracy and energy-efficiency.
\end{abstract}

\section{Introduction}

Brain-Computer Interface (BCI) design, as an emerging sub-field of Machine Learning (ML) and Human-Computer Interaction (HCI), has made significant progress in recent years. It is also emerging as a typical application within the context of artificial IoT~\cite{Zhang2010,Wu2020,Qin2018}. In general, BCI systems reply on a head-worn device to collect electroencephalography (EEG) signals and interpret them into various user attentions. Based on this technology, many experimental BCI systems have been proposed in different scenarios. For example, \citeauthor{Akram2015} \cite{Akram2015} studied how to extract a user's EEG signal to simulate a mouse click action on a PC. \citeauthor{Mauss2009} \cite{Mauss2009} took a step further where they designed a system to extract that signal from one participant and then transmit it into another participant's mind to study whether it can influence participants' video game playing behavior. \citeauthor{Pinheiro2016} \cite{Pinheiro2016} dived into a different domain where they aimed to design a BCI system that allowed patients to control a robot in the healthcare domain.

The advancements of EEG-based BCI also attribute to the powerful neural network architectures released in recent years. Nowadays, researchers can use advanced neural network based models, as opposed to the early-days regression models, to interpret the EEG signals~\cite{Gudmundsson2007}. This approach works extremely well with recurrent neural network model architectures (RNN) and its derived Long Short-Term Memory (LSTM) architecture, as the EEG signal is a chronological sequence of data.

\begin{figure}[htb]
	\renewcommand\arraystretch{1.1}
	\begin{center}
		\begin{tabular}{cc}
			\includegraphics[width=0.44\linewidth]{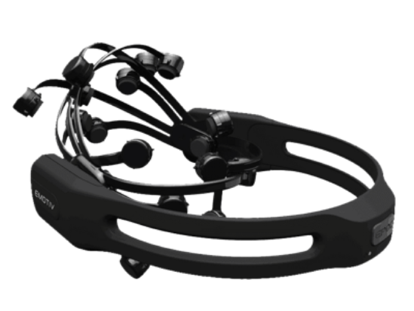} &
			\includegraphics[width=0.44\linewidth]{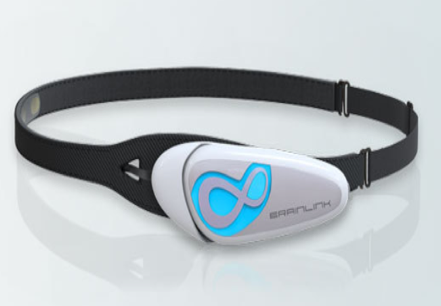}\\
			{\scriptsize (a) High-cost Emotiv EPOC} & {\scriptsize (b) Low-cost Brainlink}
		\end{tabular}
	\end{center}
	\caption{Commercially available, consumer-grade EEG collection devices.}
	\label{fig_devices}
\end{figure}

However, these existing systems suffer from a common drawback that most of them are experimental prototypes, or they were developed for institutional users (e.g., hospitals and governments~\cite{Williams2015}). Thus, the hardware cost is rather expensive, which inhibits the wide adoption in people's daily use scenarios. EEG collection equipments have different prices.
As shown in Figure~\ref{fig_devices}, EMOTIV EPOC+ 14 Channel Mobile EEG costs \$799.00 whereas Brainlink costs \$99. The usage of such EEG collection equipments for ordinary users is often limited by the price. Another drawback of the BCI-controlled robots is that they only allow users to perform one action, e.g., using a specific pattern of signals to move the robot forward~\cite{Nguyen2015}.

Our key motivation is to provide an accessible end-to-end solution for the general public users. Many people having difficulty on walking or other movements, e.g., the elderly live alone at home and they have needs to get many ``simple'' housework done, but those simple tasks (e.g., pick up a remote from the floor) are not simple for them anymore. We aim to design a solution for these users to use only their brains to control a robot performing these actions. This solution has to be low-cost, energy efficient, performing more-than-one categories of tasks, and at a relatively high accuracy. To ensure we are getting the users' needs correctly, we adopt participatory design~\cite{Muller2003} approach to invite targeted users to be part of the iterative design process. We spend days visiting the elderly care facilities and some of the houses where an elderly live alone. Through the analysis of observation note, interview, and participatory design session, we evolve our initial design (e.g., a NAO robot) into the final design (e.g., a TX2-based robot) after multiple iterations.

\subsection{Contributions}
To tackle these challenges, we propose and develop an end-to-end solution that enables fine brain-robot interaction (BRI) through embedded learning of coarse EEG signals from the low-cost devices, namely DeepBrain, so that our targeted users can mindly command and control a robot to perform some basic household tasks. The main contributions of this work are as follows:
\begin{itemize}
\item On technical aspect, we first present a Stacked Long short-term memory (Stacked LSTM) structure with specific pre-processing techniques to handle the time-dependency of EEG signals and their classification. Then we propose personalized design in the DeepBrain to capture multiple features and achieve accurate recognition of individual EEG signals by enhancing the signal interpretation of Stacked LSTM with attention mechanism. Thus, our DeepBrain can present comprehensive capabilities to process time-dependency and personal features of EEG signals at the same time.

\item On experimental aspect, we collect two datasets (one dataset in a quiet environment and the other one is in a practical but noisy environment). These datasets are from different gender- and age-groups (3 males and 3 females, span between 40 to 70) to illustrate the performance of our method. We compared our DeepBrain approach with the current state-of-the-art works. The experimental results demonstrate that our method outperforms other methods on run-time speed, accuracy and energy-efficiency.
\end{itemize}

\section{Related Work}
This section presents current research on Wearable EEG devices and EEG processing using recurrent neural networks.

\subsection{Wearable EEG Devices}

Since emotions have many tracks inside and outside our body, kinds of methods have been adopted for constructing emotion recognition models, such as facial expressions, voices, and so on~\cite{Calvo2010}. Among these approaches, EEG-based methods are measured to be promising approaches for emotion recognition. Many findings in neuroscience support that EEG allow direct assessment for the "inner" states of users~\cite{Jenke2014}. However, most of these researches link much with wet electrodes (some with dozens of electrodes). Except the time costs and high price for placing the electrodes, the unrelated channels may interfuse noise in the system, which can affect the performance of the system badly. The HCI community invokes user-friendly usage for effective brain-computer interactions. With the rapid development of wearable devices and dry electrode techniques~\cite{Huang2015}, it is possible to develop wearable EEG application devices. For instance, a language or hand disabled person wearing such a device could show his or her emotions to service robot if the device detects that he or she is in certain kind of emotional state. As we all know, an emotional recognition EEG device with easy installation is popular in HCI. In order to realize this idea, in our paper, we apply a relatively lower number of electrodes for EEG collection, the EEG collection device is called Brainlink, which has two non-invasive dry electrodes in the forehead position. When the user is in a different state (focused or relaxed), the Brainlink will display different colors of breathing lights. And then perform emotion recognition on the collected EEG data through the system described above~\cite{WeiLongZheng2018}.

\subsection{EEG Processing Using Recurrent Neural Networks}
When a neural network is used to EEG data, the connection between the data at the time before and after can be finished by manually building a sliding window to deal with the EEG time series data. Deep neural networks~\cite{LeCun2015} are applied to classify EEG data. LSTM~\cite{Schmidhuber1997} are recurrent neural networks (RNNs) equipped with a special gating mechanism that controls access to memory cells. Long short-term memory cells controlled by gates allow information to pass unmodified over many time steps. Since the gates can prevent the rest of the network from modifying the contents of the memory cells for multiple time steps, LSTM networks preserve signals and propagate errors for much longer than ordinary RNNs. Through independently reading, writing and erasing contents from the memory cells, the gates can also be trained to deal with the selection of input signals and negligence of other parts. Stacked LSTM~\cite{Graves2013} consists of LSTM unit connections along depth dimension. It benefits to store and generate longer range patterns and is much robuster. The attention LSTM cells are present along the sequential computation of each LSTM network. However, they are not present in the vertical computation from one layer to the next. Multidimensional LSTM~\cite{Schmidhuber2007} replaces a single recurrent connection with many recurrent connections, so that it can deal with multi-dimensional tasks such as images and videos.

Recently, Zhang et al.~\cite{Zhang2017} present a brain typing system for converting user's thoughts to texts via deep feature learning of EEG signals. Classifier used in their systems achieves the accuracy of 95.53\% on multivariate classification. their latest research work in~\cite{Lina2018} builds a universal EEG-based identification model which achieves the accuracy of 99.9\% without any actual system. However, in the above research approaches, some methods collect only one-channel EEG from the frontal cortex with the device, which is limited to calculate the emotion index. Some approaches focus on more channels EEG from the frontal cortex with more advanced equipment. Although the advanced devices can bring us more diverse EEG data, as is known to all, every additional electrode costs a lot of money. Therefore, we may spend thousands of dollars on data collection devices finally, which is not good for us to design a complete and feasible BCI system. Few studies attempt to build a feasible, high precision, civilian and easily deployable EEG-based emotion recognition system. However, this approach is only applicable to short-term dependent time series data. Our approach uses an enhanced LSTM prediction model. This model introduces the explicit internal LSTM unit structure and frequently updates the internal state values while acquiring input data at each time point, which guarantee that the EEG data before and after the time point can hold a powerful connection.

\section{System Overview}
  Our solution consists of two subsystems: EEG signal collection and pre-processing module, and neural-network-based EEG signal interpreter.
   The main goal of our method is to design a deep learning model that classifies the user's emotion status with raw EEG signal generated by our low-cost equipment in real-time.

  \begin{figure}[htb]
    \centering
		\includegraphics[width=.95\linewidth]{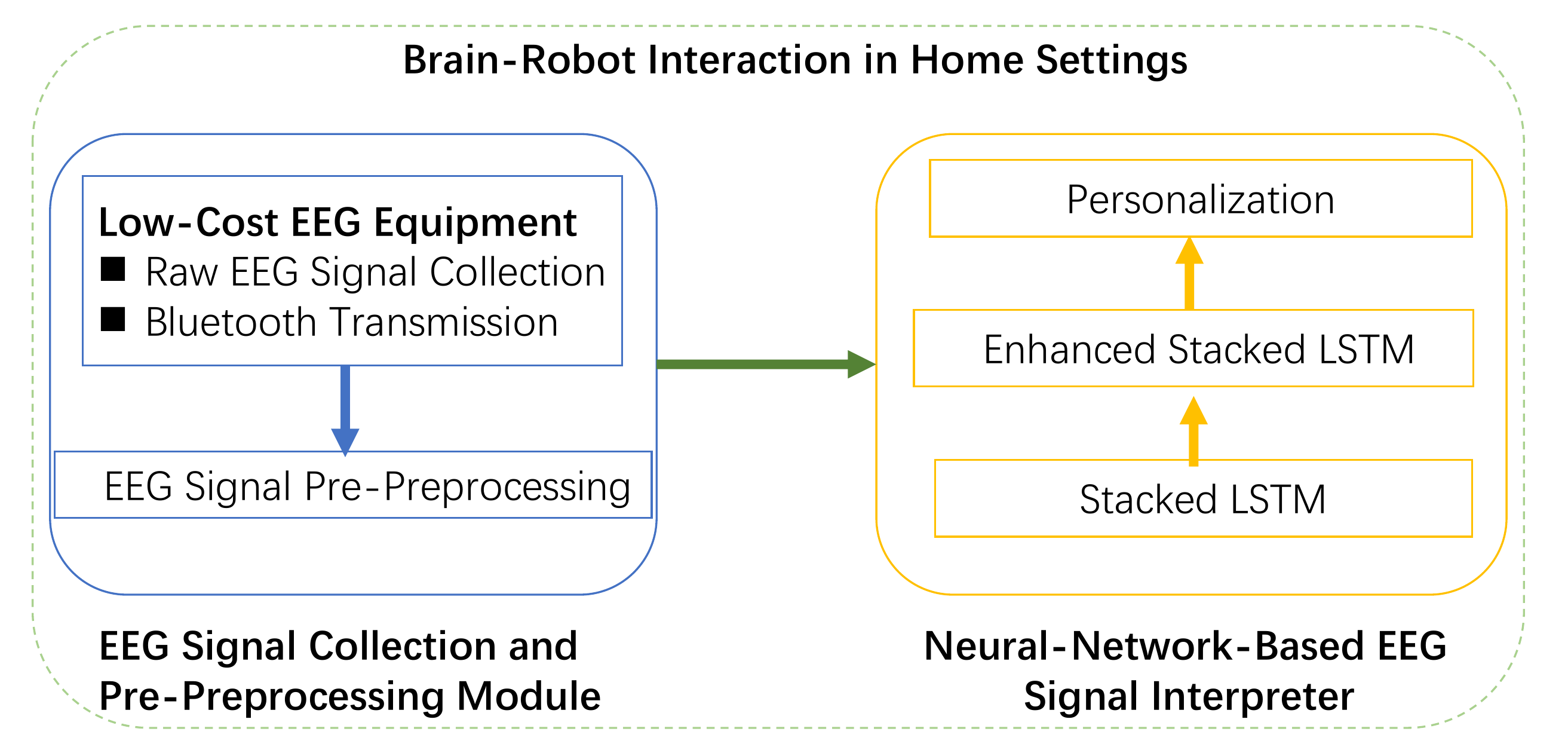}
	\caption{System architecture diagram.}
	\label{fig:system_arch}
\end{figure}

  In the EEG signal pre-processing section, we use an off-the-shelf headset equipment to collect raw EEG signal; then we feed the signal into a EEG signal pre-processing algorithm for multi-classification.

  The neural-network-based EEG signal interpreter module  read in the processed EEG signals and translate it into one of the four status with the help of LSTM network architecture. First, we use stacked LSTM to process the long-time dependency. Second, we propose an attention-based enhanced stacked LSTM to capture user's EEG signal status. It is worth mentioning that we also incorporate a personalization step in this module so that we can achieve a more accurate predication results after a simple fine-tuning step.

\section{LSTM-Based Method Processing EEG Signal}
We propose a hybrid deep learning model to interpret the raw EEG signals. In this part, we first summarize the pre-processing step for EEG signals. Then, we outline the proposed LSTM-Based method and its various components. At the end, we introduce the technical details Brain-robot interaction system in subsequent subsections.

\subsection{EEG Signal Collection and Pre-processing Module}
\begin{figure}[htb]
	\renewcommand\arraystretch{1.1}
	\begin{center}
		\begin{tabular}{cc}
			\includegraphics[width=0.45\columnwidth]{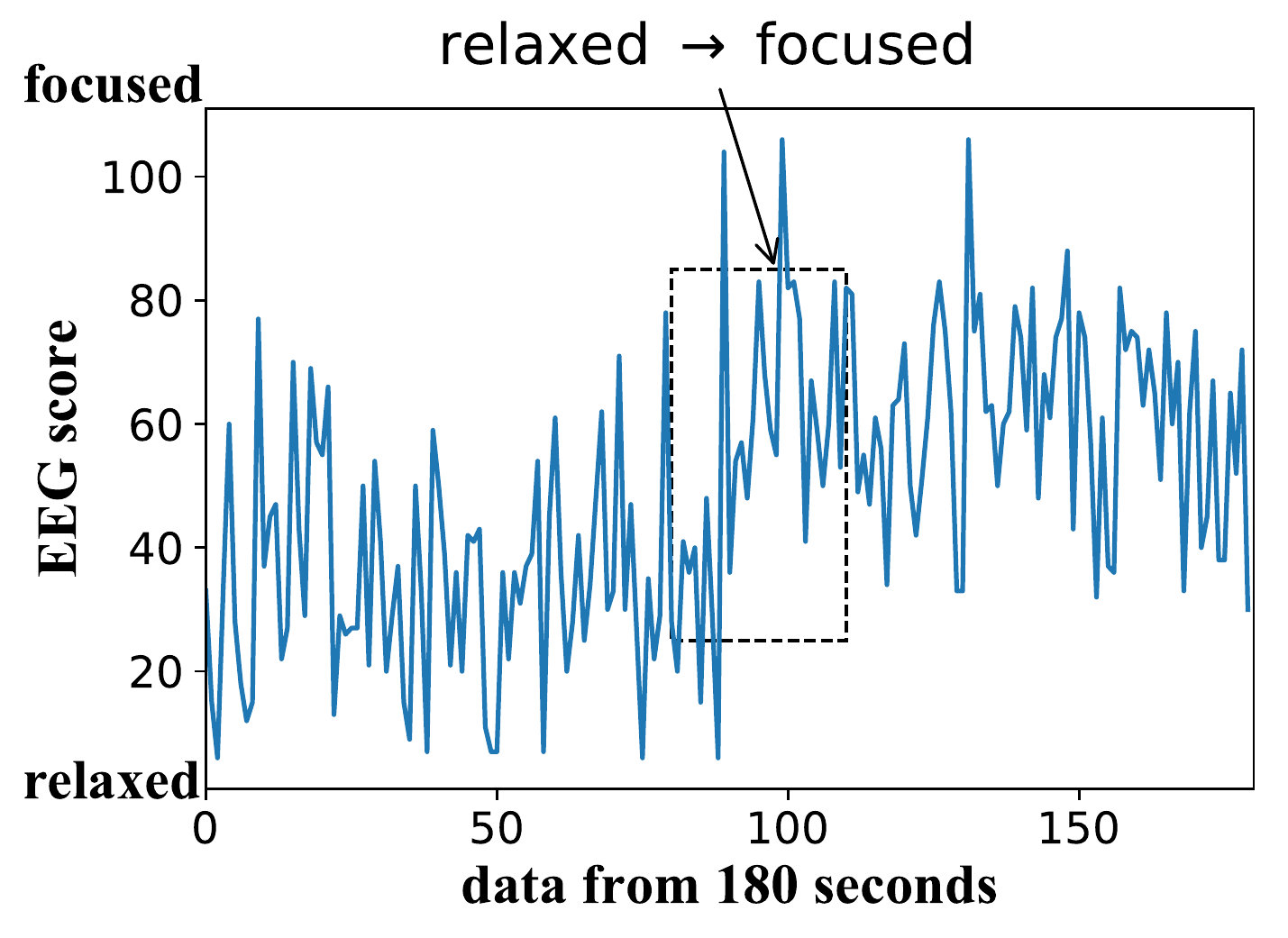} &
			\includegraphics[width=0.45\columnwidth]{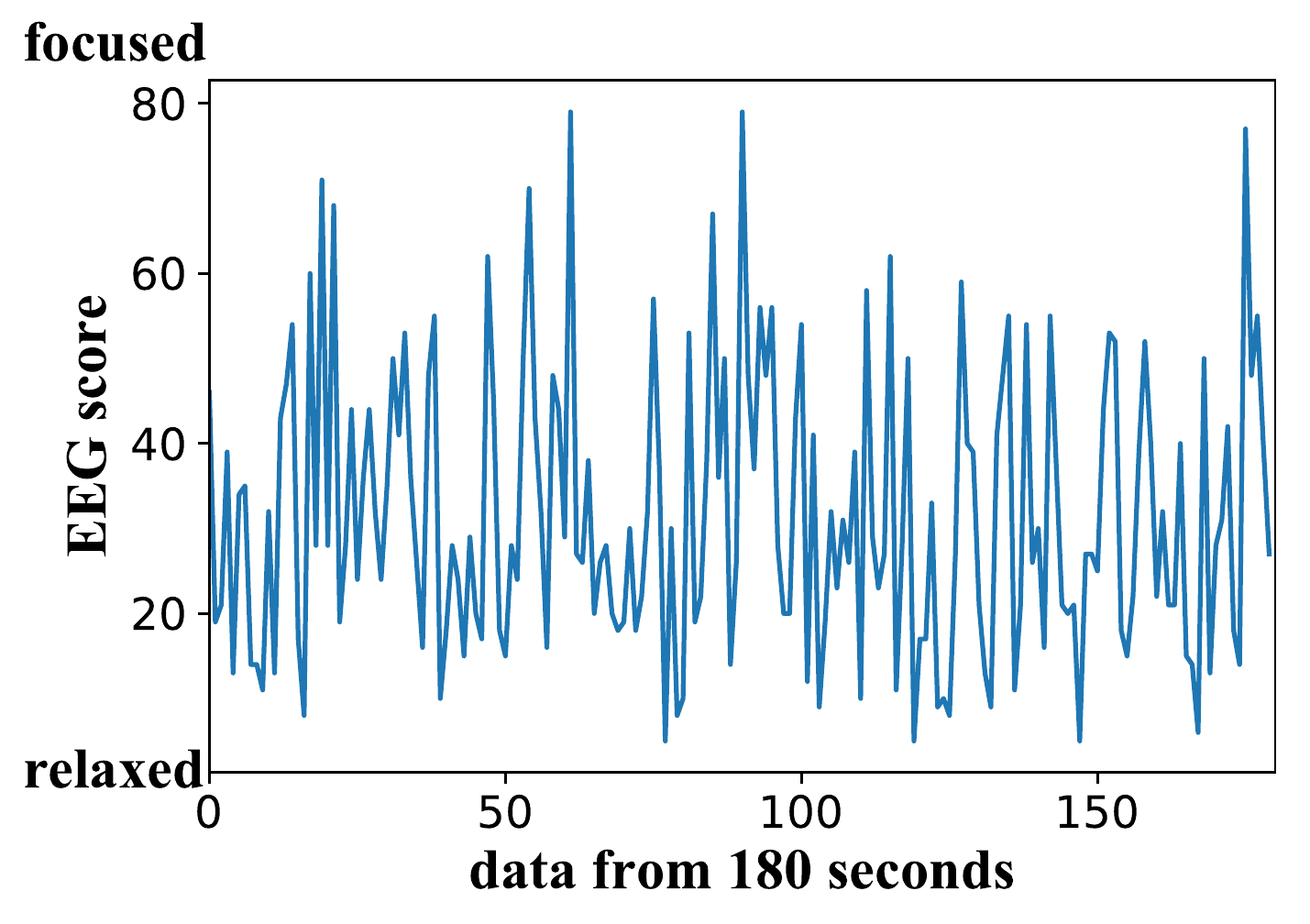} \\
			{\scriptsize (a) EEG signals from relaxed to focused} & {\scriptsize (b) EEG signals keep relaxed. }\\
			\includegraphics[width=0.45\columnwidth]{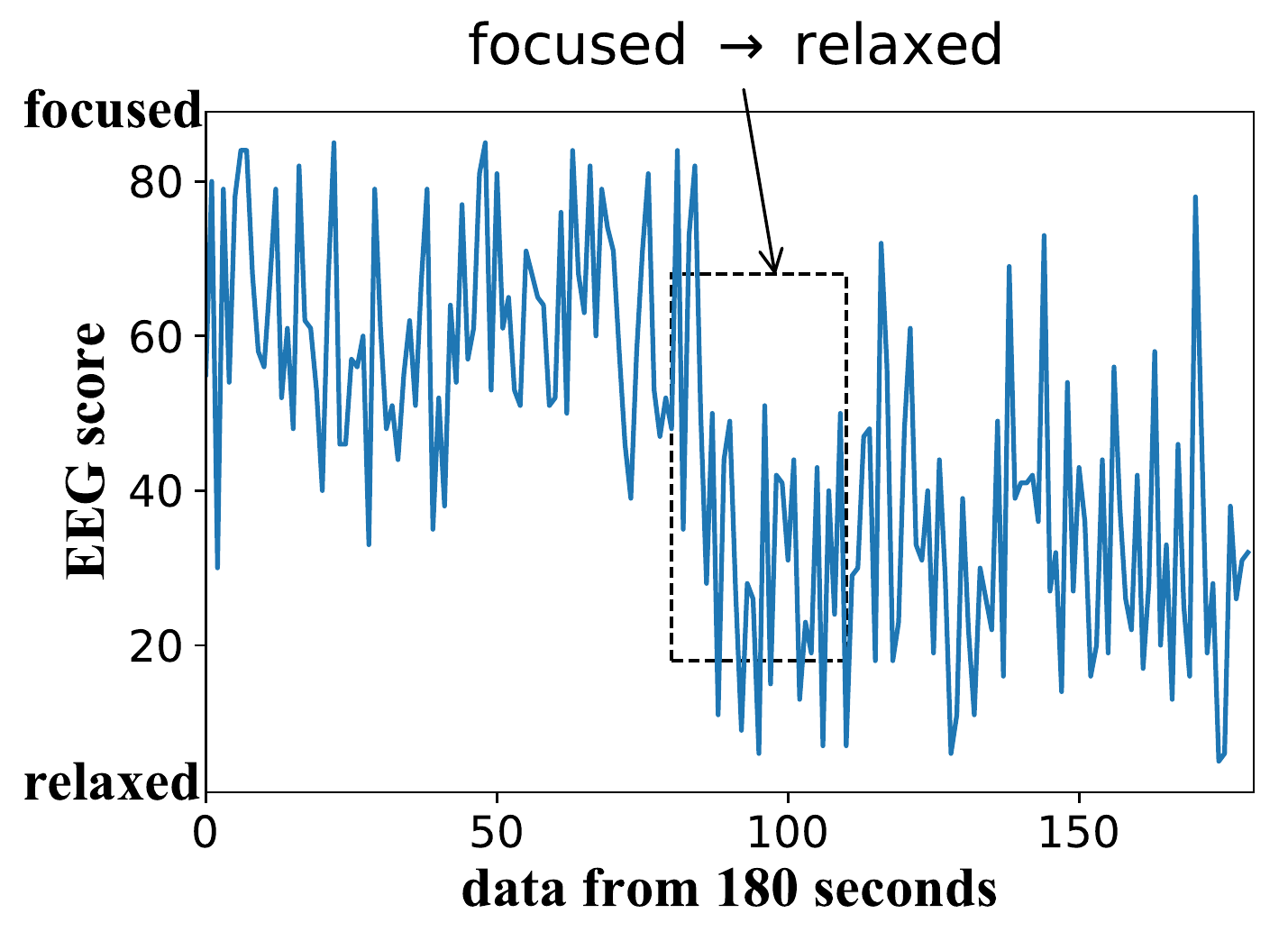} &
			\includegraphics[width=0.45\columnwidth]{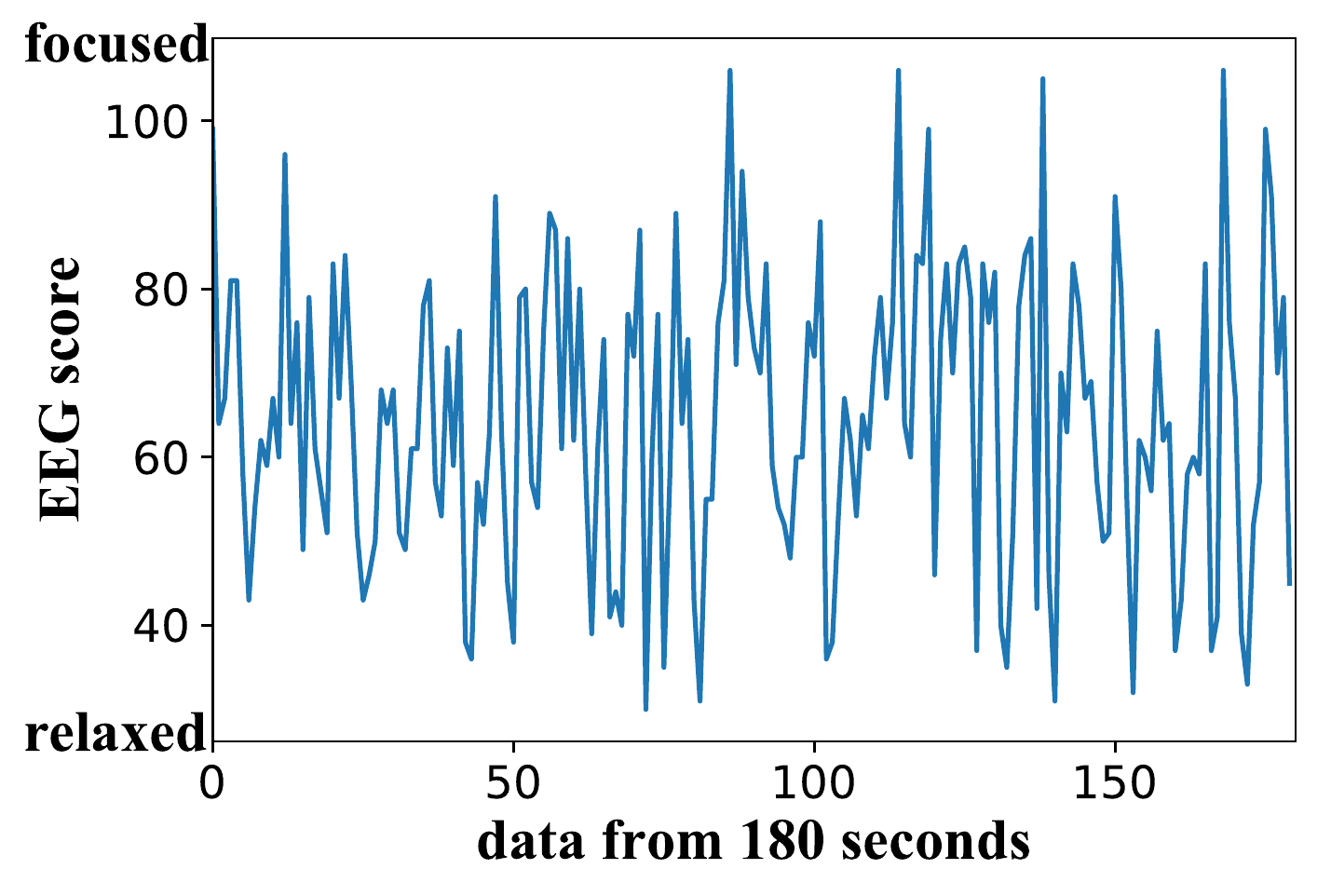}\\
			{\scriptsize (c) EEG signals from focused to relaxed.} & {\scriptsize (d) EEG signals keep focused.}\\
		\end{tabular}
	\end{center}
	\caption{Unpreprocessed EEG signals describe four states.}
	\label{fig:eeg_origin}
\end{figure}
This paper uses low-cost EEG data collection devices. Although the device is resistant to noise from multiple signal channels from users, it is occasionally affected by external environments such as weather and sound. Therefore, there are sometimes unusual data points in the data set. To improve the accuracy and stability of the results, it is necessary to perform data preprocessing. The unprepared EEG data collects in 180 seconds is shown in Figure~\ref{fig:eeg_origin}. It can be seen that the data is quite dense and confusing. If there is no appropriate means for data preprocessing, it will train the model and forecasting brings big challenges. In Figure~\ref{fig:eeg_origin}, the low-cost EEG data collection device provides a score value that reflects the decimal representation of the user's brain emotional state, which is a numerical representation of the above-mentioned EEG pattern. When the user's EEG signal value is high, it indicates that the brain has a high probability of being in the Beta or Gamma pattern. Conversely, when the user's value is low, it indicates that the brain has a high probability of being in the Delta or Theta pattern. As for the user's medium-conscious Alpha pattern, the low-cost EEG data collection device will give corresponding numerical values according to different users (male or female).

\begin{figure}[htb]
	\renewcommand\arraystretch{1.1}
	\begin{center}
		\begin{tabular}{cc}
			\includegraphics[width=0.49\columnwidth]{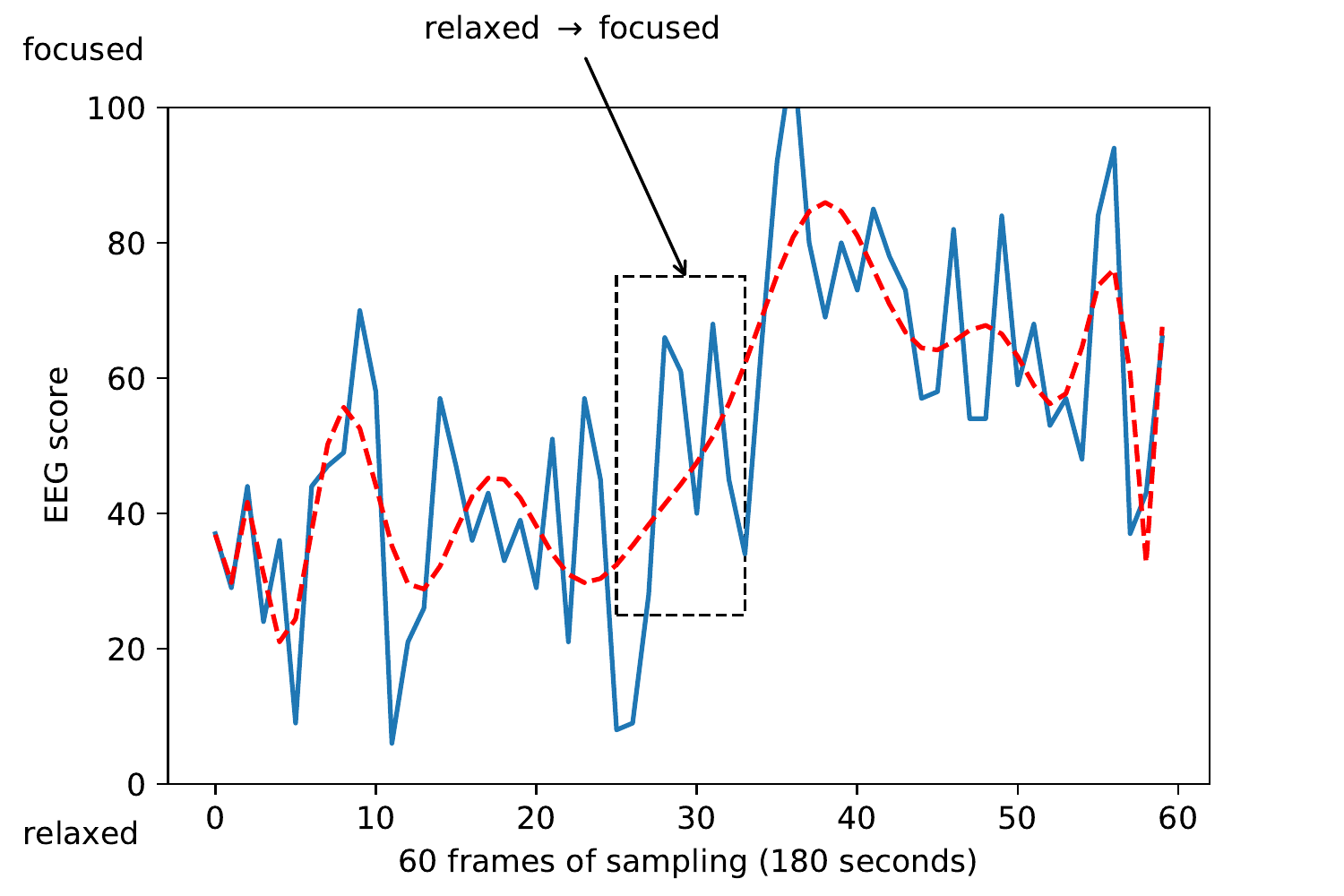} &
			\includegraphics[width=0.49\columnwidth]{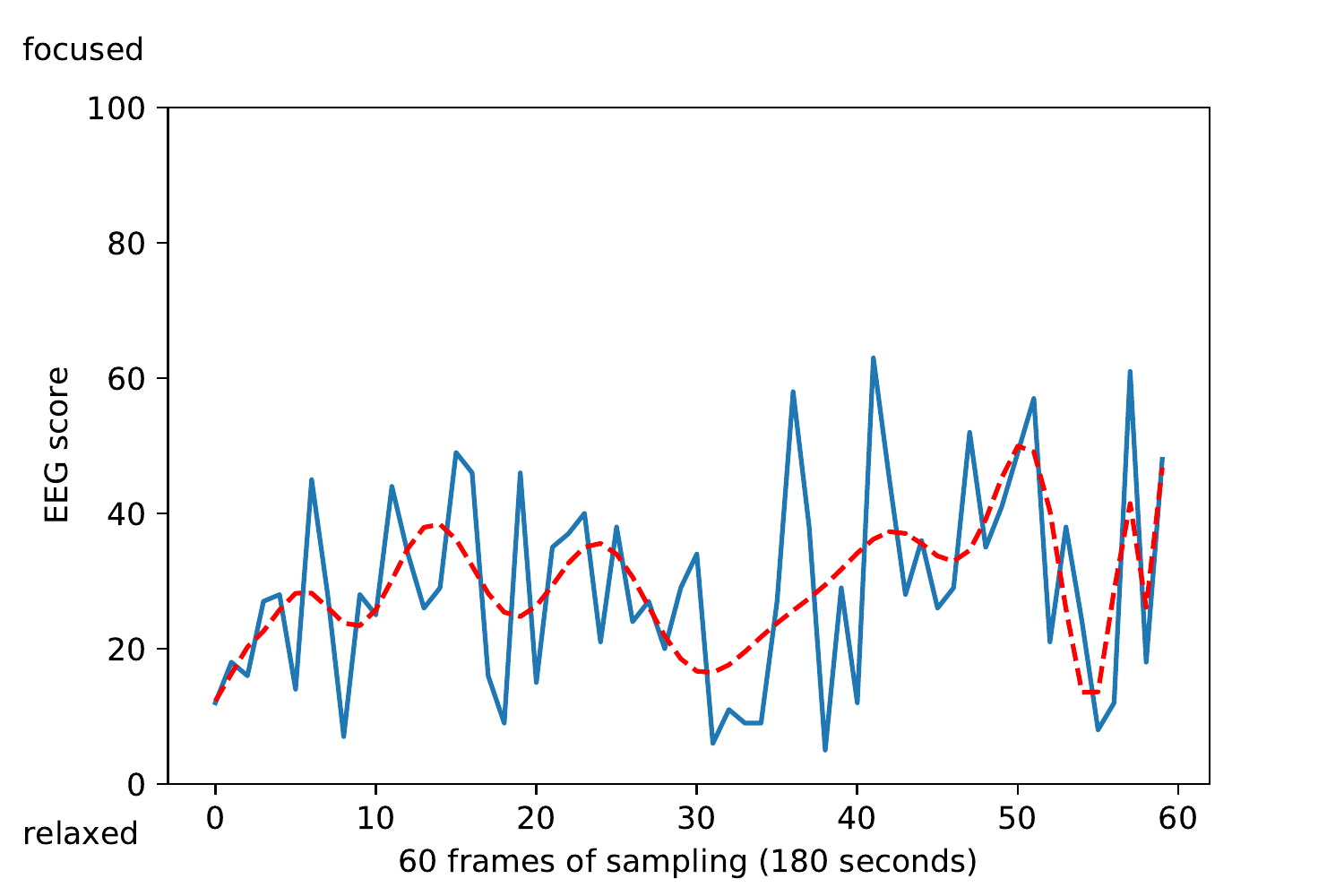} \\
			{\scriptsize (a) The status from relaxed to focused.} & {\scriptsize (b) The status keep relaxed. }\\
			\includegraphics[width=0.49\columnwidth]{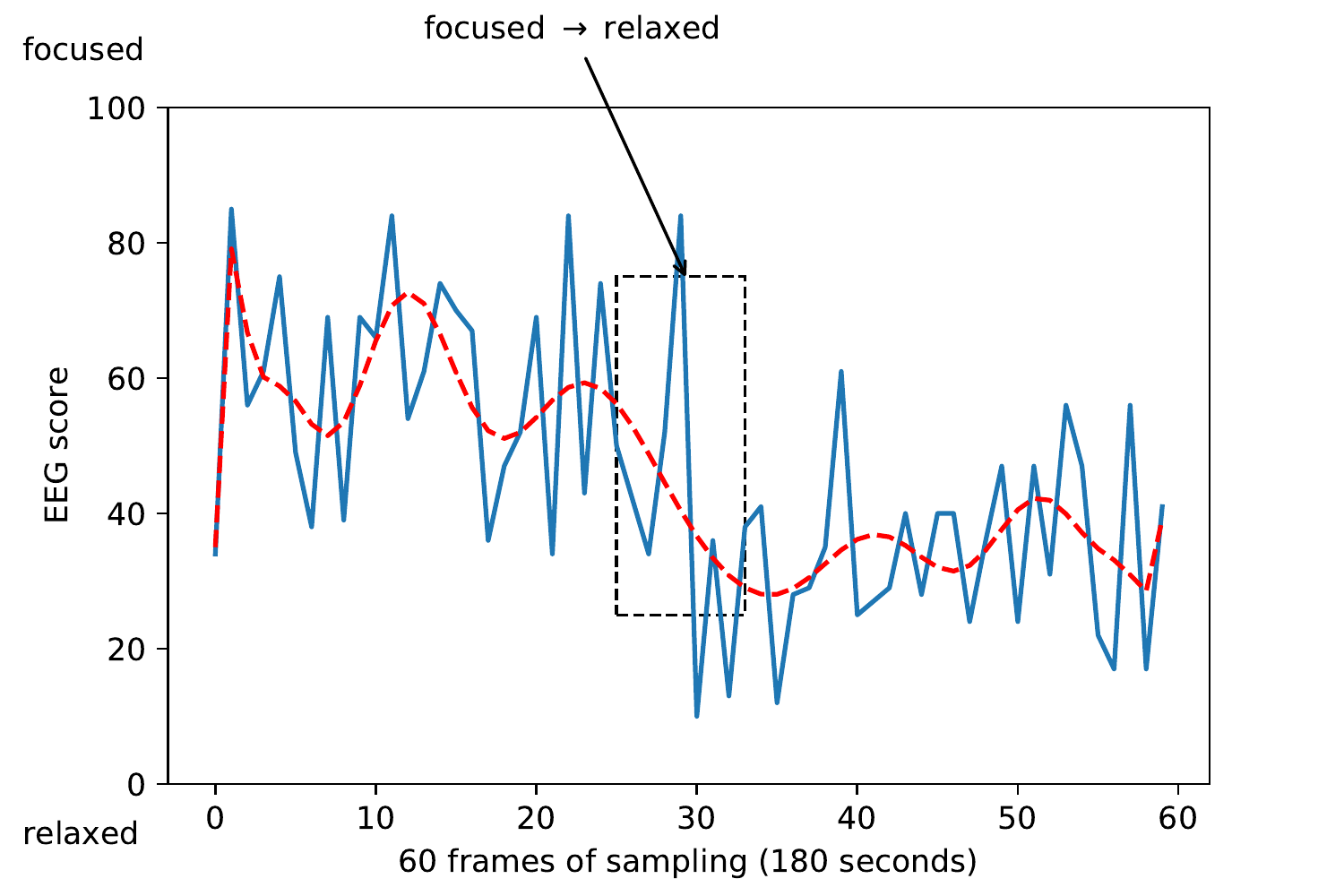} &
			\includegraphics[width=0.49\columnwidth]{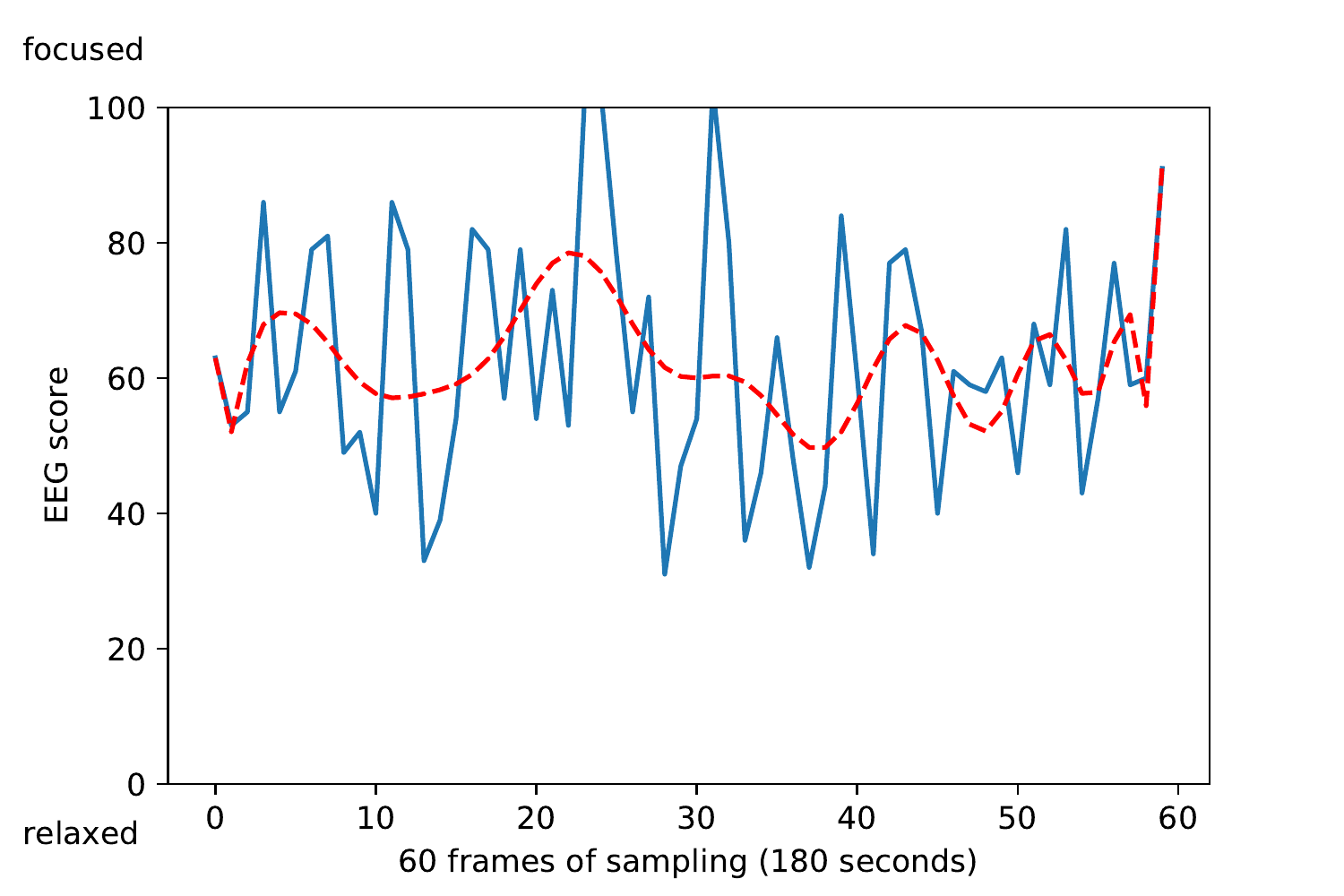}\\
            {\scriptsize (c) The status from focused to relaxed.} & {\scriptsize (d) The status keep focused.}\\
		\end{tabular}
	\end{center}
	\caption{EEG signal sessions as an example to illustrate four status: focused, relaxed, focused$\rightarrow$relaxed, and relaxed$\rightarrow$focused.}
	\label{fig:Original}
\end{figure}
\begin{figure}[htb]
	\renewcommand\arraystretch{1.1}
	\begin{center}
		\begin{tabular}{cc}
			\includegraphics[width=0.47\columnwidth]{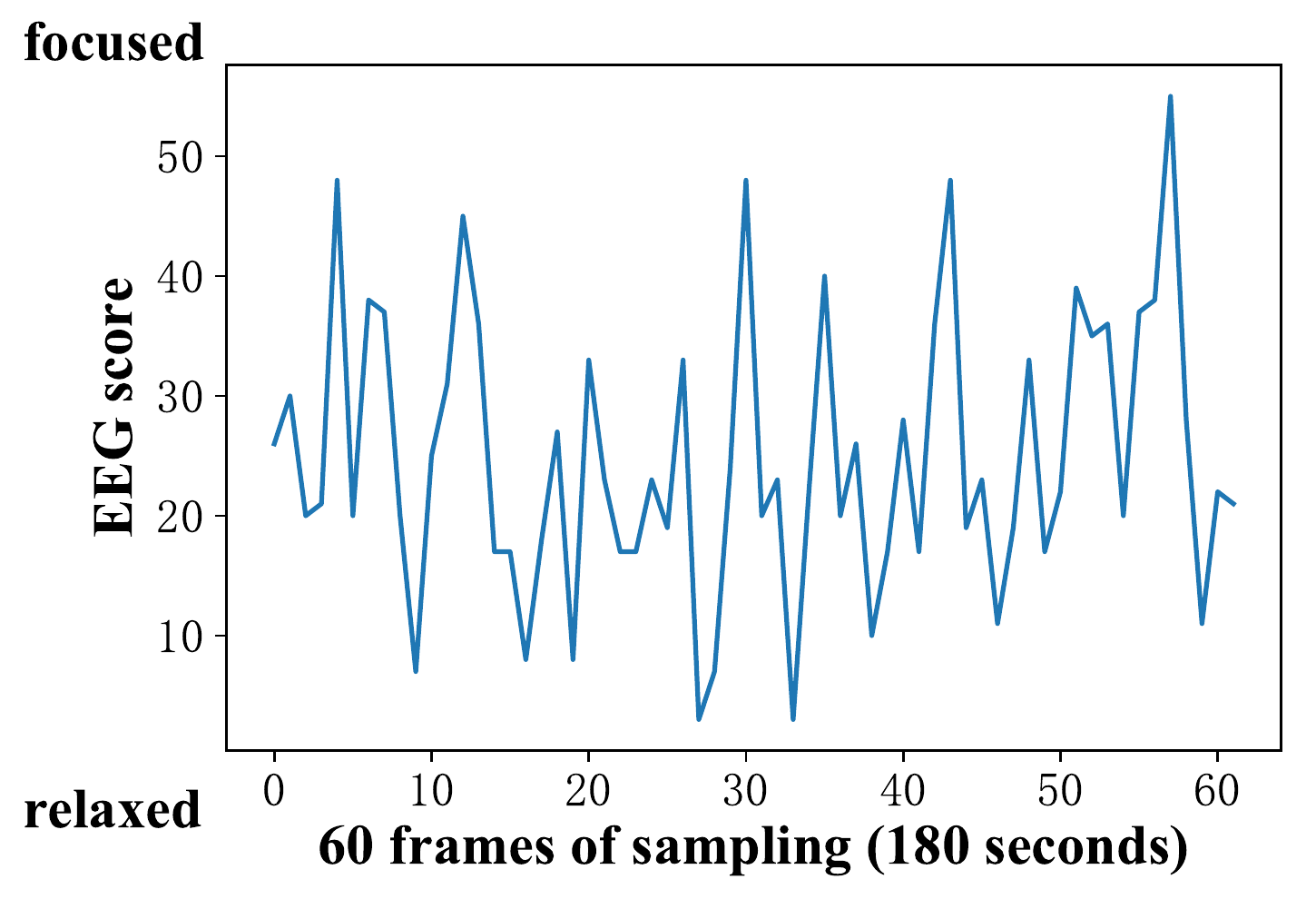} &
			\includegraphics[width=0.47\columnwidth]{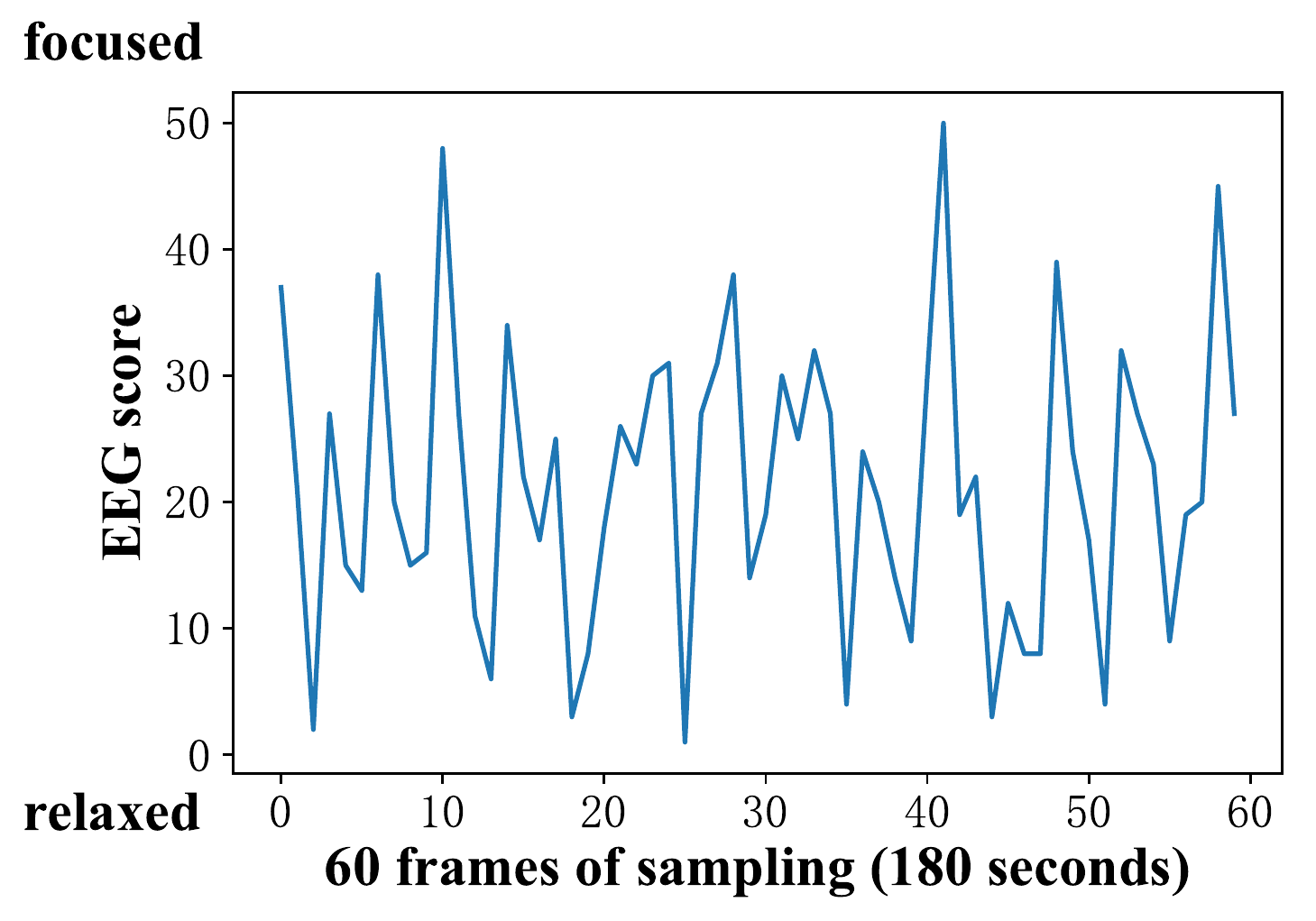} \\
			{\scriptsize (a) Alpha pattern for male} & {\scriptsize (b) Alpha pattern for female}
		\end{tabular}
	\end{center}
	\caption{The difference of EEG signal Alpha pattern between male and female in the same age.}
	\label{fig:mid_signal}
\end{figure}
Furthermore, we use low-cost equipment to collect EEG data. Compared to high-cost equipments, such as Emotiv Epoc+11 headset, which detects 5 types of patterns (Delta, Theta, Alpha, Beta and Gamma), Our EEG equipment can only detect the two types of patterns, which is high and low. The signal is low value, and for the middle Alpha pattern, according to different people (such as male and female). The signal value is divided into focused or relaxed. Therefore, based on the changes in the two states, we divide the local datasets' label into four categories. For Alpha pattern in the middle of Delta to Gamma, we need to divide it into relaxed or focused based on the signal value from male or female.

First of all, our EEG collection device gives decimal scores that reflect the user's brain emotion status, as shown in the Figure~\ref{fig:Original}. Our EEG equipment is resistant to the noises from multiple channels of signals from the user, it can occasionally be affected by the external environment, such as weather and sound. Thus, sometimes there would be an outlier data point in the dataset. We apply statistic methods to identify the outlier and discard them. And we use the average score of the data point before and after it (as all the data is a time series sequence) to represent it.

As described at the beginning of the article, we properly encode the two types of collected data , which achieves the expected four classification effects. This article uses one of the typical encoding methods named one-hot encoding for label processing, which can effectively avoid the model abrupt problem caused by the logarithmic calculation of the model during training. In the initial stage of multi-classification coding operation, the experimental results obtained by our experiments are always in a low numerical range. We think that the situation may be caused by two factors, one is that there has a problem with the multi-class model architecture, the other one is that the entered data cannot be classified correctly. Through the follow-up experiments, we exclude the first case and finally find that the problem occurs in the second case, because the multi-class model structure has no problem. We list the problems as follows: 1) There is a specific correspondence between the classification data and the label. 2) There may be some similarity in the features of the classification data extraction. 3) Classification data cannot be classified. 4) The categorical data can be a feature combination (A\&B) but not a feature (A1\&A2). The experimental data that we collect conflict with case 3) above, which results in the model not being able to classify the data normally. Based on this, we give some additional explanations about case 3). If we now want to identify the ear characteristics of an animal, such as cat ears and dog ears, we can get the correct classification result by training the appropriate model. However, if the characteristics of dog ears and cat ears are each taken in half and combined into a new feature, the model would not correctly classify the characteristics of the combination during the operation. For the middle Alpha pattern, we divide it into relaxed or focused based on the EEG signal value of the male or female user. As shown in the Figure~\ref{fig:mid_signal}, the intermediate state of the EEG data is higher for male, even in the relaxed state, the value is still close to 60, while the value of female is less than 50.

Therefore, the model can only judge the category by random division, which leads to the low numerical range of the experimental results. If the above-mentioned mixed condition occurs in the feature, it is not good to find a suitable function for feature classification. Taking the situation into consideration, we perform secondary separation on the mixed features, and the designed function needs to satisfy the following conditions: $F(u,v) \neq F(v,u)$ , and the regions represented by $F(u)$ and $F(v)$ are isolated at $u=v$. Here we give the mathematical formula used in the preprocessing, $Y=2A-B$, where A is the raw data, B is part of the raw data. We have re-sampled the data several times to verify the feasibility of the formula. In addition, we have considered other functions, such as $Y=A^2-B^2+AB$, where A is the raw data, B is part of the raw data, etc., but there are always various problems to reduce the accuracy of the model.

\subsection{Neural-Network-Based EEG Signal Interpreter}
We focus on learning the meanings of the user's intent signals which are 1-D vectors (collected in one time point). We express the single input EEG signal as E$_i$. Then, we feed E$_i$ to the LSTM structure for temporal feature learning. At last, according to the learned temporal features X$_t$, the result of the classification is given~\cite{Stober2015}.

\begin{figure}[htb]
	\begin{center}
		\includegraphics[width=\columnwidth]{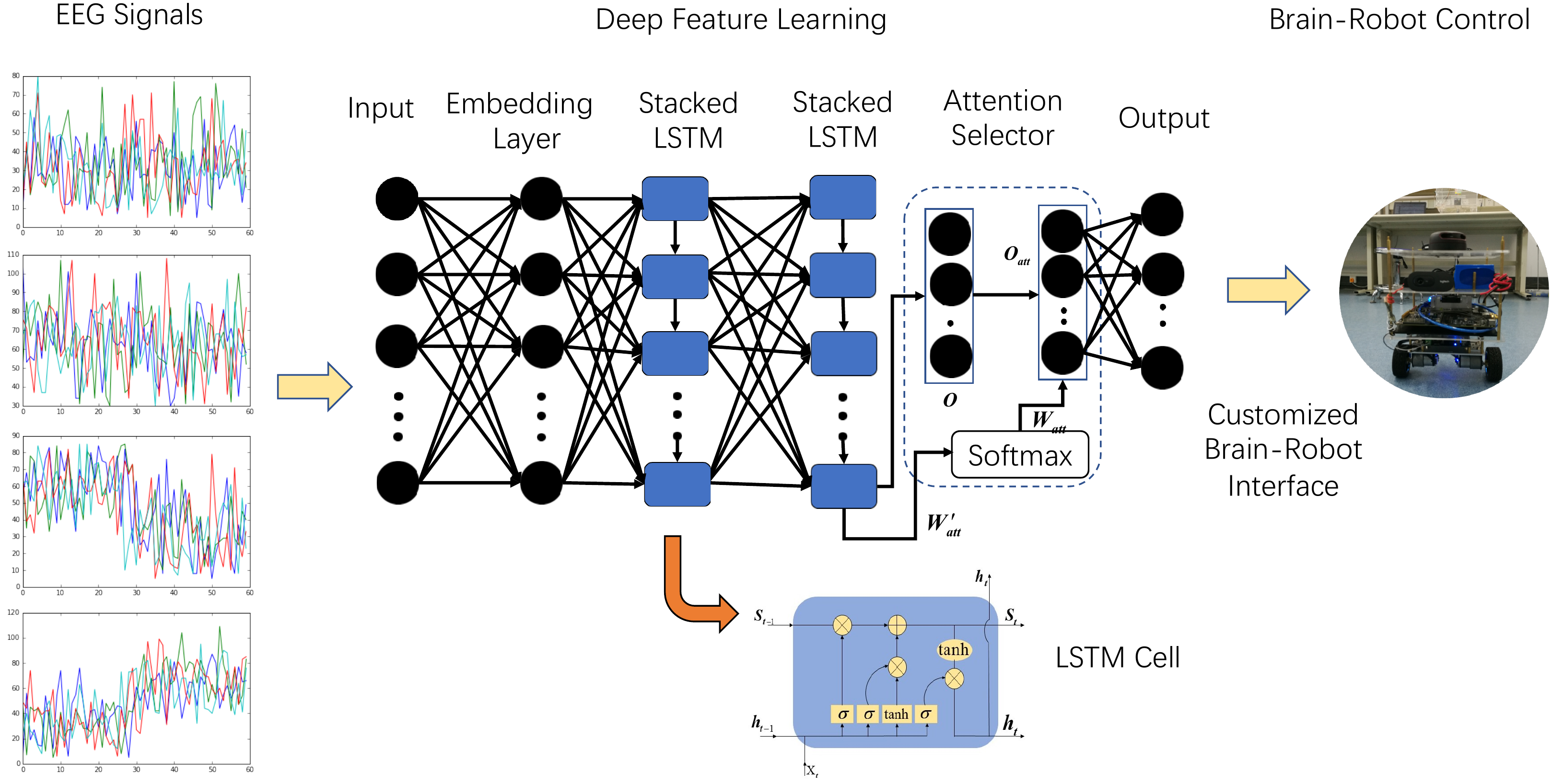}
	\end{center}
	\caption{The illustration of DeepBrain workflow and interaction operations}
	\label{fig:lstm_arch}
\end{figure}

 The central idea of our DeepBrain workflow and interaction operations is depicted in Figure~\ref{fig:lstm_arch}. The input raw EEG data is a single sample vector. We first utilize two fully connected layers as our hidden layer, and then input its value of output to the LSTM units. In addition, the arrow shows the internal structure of the LSTM layer, where $\sigma$ and $tanh$ represent the activation function. $X_t$ is the input of model. $h_t$ is the output of LSTM cell in the $t$-th time step and $h_{t-1}$ is derived in the previous sequence step. $S_t$ stands for the value of LSTM memory cell in the $t$-th time step.

For the challenges of time series data processing, we first use down sampling technique to obtain the characteristic subsequence of the original time series. Down sampling reduces the complexity of the original time series and makes it easier for learning patterns. At the same time, to speed up the convergence rate of the model, we normalize our time series data using min-max normalization, which is a linear transformation of the original data. The transformed values are mapped into the interval [0, 1].

In the temporal feature processing part, the powerful time feature extraction capability of the LSTM structure is proven. LSTM can explore the feature dependencies over time through an inner state of the network, which permits it to display trend temporal behavior. LSTM cells control the input, storage and output of data by introducing a set of gate mechanisms. As shown in the Figure~\ref{fig:lstm_arch}, the LSTM gate units receive the output of the LSTM internal units at the previous time step and the input at current time step sample. If the previous layer of the LSTM cell layer is not the input layer, its various gate units accept the output of previous layer's LSTM internal units at the current time step and the output of the LSTM internal units at the previous time step. We utilize an LSTM model that contains three components: one input layer, 2 hidden layers, and one output layer. LSTM (shown as the rectangles in the Figure~\ref{fig:lstm_arch}) cells are in the hidden layers. Assume that a batch of input EEG data contain $n_s$ (generally called batch size) EEG samples and the total input data have the 3-D shape as $[n_s, 30, 1]$.
Let the data in the $i$-th layer be denoted by $X^r_i$ = $\{X^r_{ijk} | \ j = 1, 2, \cdots, n_s, \  k=1, 2, \cdots ,K_i \}$, $X^r_i \in \mathbf{R}^{[n_s, K_i, 1]}$, where $j$ denotes the $j$-th EEG sample and $K_i$ denotes the number of dimensions in the $i$-th layer.

Assume that the weights between layer $i$ and layer $i+1$ can be denoted by $W^r_{i(i+1)} \in \mathbf{R}^{[K_i,K_{i+1}]}$ , e.g., $ W^r_{23}$ means the weight between layer 2 and layer 3. $b^r_i \in \mathbf{R}^K_i$ denotes the biases of the $i$-th layer. The calculation between the $i$-th layer data and the $i+1$-th layer data can be denoted as $$X^r_{i+1} = X^r_i * W^r_{i,i+1} + b^r_i$$

The calculation of LSTM layers are shown as follows: $$f_i = sigmoid(H(X^r_{(i-1)j},X^r_{(i)(j-1)}))$$ $$f_f = sigmoid(H(X^r_{(i-1)j},X^r_{(i)(j-1)}))$$ $$f_o = sigmoid(H(X^r_{(i-1)j},X^r_{(i)(j-1)}))$$ $$f_m = tanh(H(X^r_{(i-1)j},X^r_{(i)(j-1)}))$$ $$c_{ij} = f_f \odot c_{i(j-1)} + f_i \odot f_m$$ $$X^r_{ij} = f_o \odot tanh(c_{ij})$$
where $f_i$, $f_f$, $f_o$ and $f_m$ represent the input gate, forget gate, output gate and input modulation gate separately, and $\odot$ denotes the element-wise multiplication. $c_{ij}$ means the state (memory) in the $j$-th LSTM cell in the $i$-th layer, which is the most important part to quest the time-series relevance among EEG data samples. $H(X^r_{(i-1)j},X^r_{(i)(j-1)})$ means the operation as follows: $$X^r_{(i-1)j} \ast W + X^r_{(i)(j-1)} \ast W' + b $$ where $W$, $W'$ and $b$ mean the corresponding weights and biases. Subsequently, we use the Back Propagation Through Time (BPTT) algorithm to train our designed model. Finally, we get the designed model prediction results and use softmax crossentropy as the loss function. The loss function is optimized by the Adam optimizer~\cite{Kingma2014} with a learning
rate of $10^{-4}$ and a minibatch size of 64.

\subsection{Attention-based Enhanced Stacked LSTM}

In addition to the proposed Stacked LSTM structure that can handle the time-dependency of EEG signals, we are aware that different people might have various EEG signal patterns and this aspect need to be carefully tackled with extended design on the Stacked LSTM. In order to present personalized solution in DeepBrain, we enhance the representation of Stacked LSTM with attention mechanism. The attention-based enhanced Stacked LSTM can facilitate the DeepBrain to learn specific features of different people, and tune our system to achieve accurate recognition of individual EEG signals.

\begin{figure}[htb]
	\begin{center}
		\includegraphics[width=0.75\columnwidth]{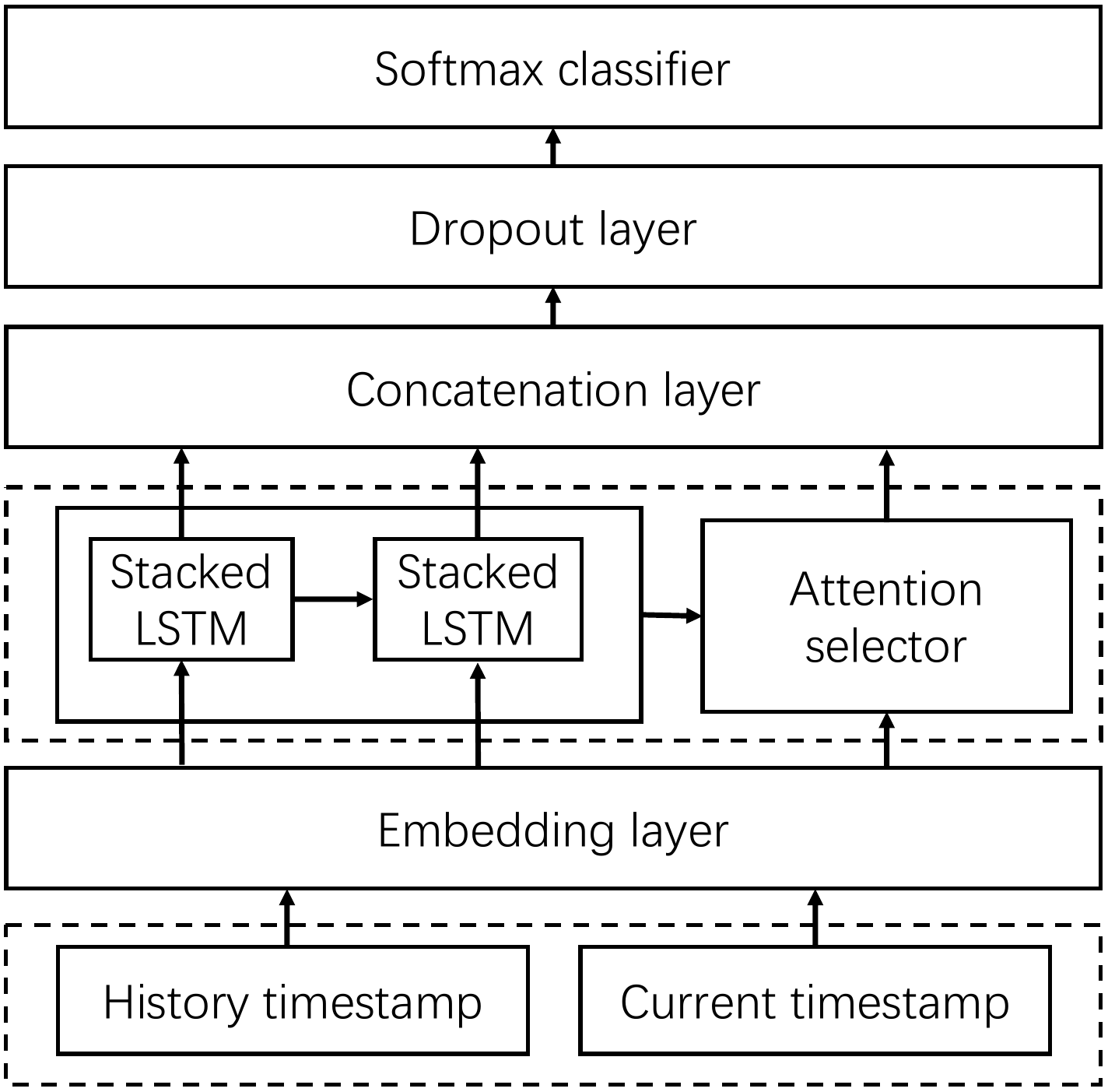}
	\end{center}
	\caption{Attention-based enhanced stacked LSTM.}
	\label{fig_attention}
\end{figure}

Figure~\ref{fig_attention} depicts the overall architecture of attention-based enhanced stacked LSTM. We use history timestamp to predict the current timestamp. The embedding layer is the first layer of attention-based enhanced stacked LSTM. And then, we use stacked LSTM to capture the long-dependency. After that, in order to customize the EEG recognition, we enhance the representation of stacked LSTM with attention selector. The attention selector accepts the final LSTM cell's input value as the attention weights $W_{att}^{\prime}$ which is measured by the operation $$ W_{att}^{\prime}=P^{\prime}\left(c^r_{i(j-1)},X^r_{i(j-1)},X^r_{(i-1)j}\right)$$ where $c^r_{i(j-1)}$ denotes the hidden state of the $(j-1)$-th LSTM cell. The operation $P^{\prime}(*)$ is similar with the calculation process of the LSTM structure and calculates the normalized attention weights $W_{att}$ as
$$W_{att}={softmax}\left(W_{att}^{\prime}\right)$$ Then, the concatenation layer is used to connect the output of Stacked LSTM and attention selector. The dropout layer is a regularization technique to improve the generalization of EEG signal recognition model. The final softmax layer is to classify four EEG signal patterns.

\section{Experiments and Results}
In this section, we use the collected local dataset to evaluate the designed deep learning model.

\subsection{Evaluation Metrics}
We use six metrics to comprehensively evaluate the performance of our model: Accuracy, Precision, Recall, F1, ROC (Receiver Operating Characteristic), AUC (Area Under Curve), TPR (True Positive Rate), and FPR (False Positive Rate). These metrics have been widely used to assess machine learning algorithms. The details of the metrics are listed following:
\begin{align*}
  Accuracy &= \tfrac{TP + TN}{TP + FN + FP + TN} & F1 &= \tfrac{2 \cdot Precision \cdot Recall}{Precision + Recall} \\
  Precision &= \tfrac{TP}{TP + FP} &  TPR &= \tfrac{TP}{TP + FN} \\
  Recall &= \tfrac{TP}{TP + FN} & FPR &= \tfrac{FP}{FP + TN}
\end{align*}

In general, accuracy is used to judge a model whose goal is to classify, meanwhile since our goal is to identify the EEG data category, precision and recall are also important metrics for evaluating our model. Precision rate is mainly used to judge whether the classifier can correctly get classification results, that is to say, it mainly focuses on identifying abnormal samples. Furthermore, recall rate mainly evaluates whether the classifier can identify all abnormal samples. $F_\beta$ score is a combination of the previous two metrics, if $\beta$ is less than 1, it represents that the recall rate is more important. On the contrary, the precision rate has a greater impact on the model quality assessment. $F1$ score is used as a general overview of the performance about the algorithm. ROC is a graph composed of a False positive rate (horizontal axis) and a True positive rate (vertical axis). We can obtain different $TPR/FPR$  pairs by adjusting the classifier's classification threshold. These data pairs are ROC data points, and they can intuitively reflect the advantages of the classifier. AUC is the area under the ROC curve, and it reflects the performance of the classification model. The closer the AUC value is to 1, the better the classification is. Four outcomes of the classification include True Positive (TP), False Positive (FP), True Negative (TN) and False Negative (FN), as shown above. We show the ROC curve under the EEG data and analyze the ROC curves of the four models.

\subsection{Experimental Settings}
We first collect data through Adriano serial communication, and we need to install corresponding driver module on Jetson TX2. The subject is asked to wear the EEG device and control robot by mind. At the beginning, we need to compile and configure the various toolkits needed for the experiment on Jetson TX2. Specifically, we install the deep learning framework TensorFlow required for the experiment by means of source code compilation. By running TensorFlow, we can smoothly load the model we designed. We have carefully annotated the EEG data to the corresponding actions that are undertaken by the subject and been available from context. In our experiments, we choose a sum of 800 labeled EEG samples collected from 4 subjects (800 samples per subject). Each sample is a vector of 180 elements and corresponds to one channel of the EEG data. To evaluate the performance, we use several evaluation metrics such as accuracy, CPU and GPU ratio, ram footprint and so on.

\subsection{EEG Signals Analysis}
\begin{table*}[htb]
	\renewcommand\arraystretch{1.1}
	\centering
	\caption{The correlation coefficients matrix of Self and Cross}
	\label{tab:mapeone}\scalebox{0.9}{$
	\begin{tabular}{|c|c|c|c|c|c|c|}
		\hline
		Class & relaxed & relaxed$\rightarrow$focused & focused$\rightarrow$relaxed & focused & \textbf{Self} & \textbf{Cross}
		\\
		\hline
		relaxed & 0.00155 & 0.04916 & -0.00463 & 0.02890 & 0.00156 & 0.02448
		\\
		\hline
		relaxed$\rightarrow$focused & 0.04916 &	0.32900 & -0.39446 & 0.05040 & 0.32900 &	-0.09830
		\\
		\hline
		focused$\rightarrow$relaxed & -0.00463 & -0.39446 & 0.42358 & 0.03106 & 0.42358 &-0.36803
		\\
		\hline
		focused & 0.02891& 0.05040 & 0.03106 & 0.01933 & 0.01933 & 0.03679 \\
		\hline
	\end{tabular}$}
\end{table*}
Furthermore, we concisely analyze the similarities between EEG signals corresponding to different intents and quantify them using spearman correlation as shown in Table~\ref{tab:mapeone}. In order to make the machine understand human intentions better, we present two similarities used in our experiment, inter-class similarity and extra-class similarity. The inter-class similarity means the similarity of EEG signals within the same meaning. We randomly choose several EEG data samples from the same intent and calculate the spearman correlation coefficient respectively. The inter-class similarity is measured as the average of spearman correlation coefficients of all samples. Likewise, extra-class similarity indicates the correlation coefficient between different EEG categories. We estimate the correlation coefficients matrix for each subject and then calculate the average matrix. Table~\ref{tab:mapeone} shows the correlation coefficients matrix and the relevant statistical extra-similarity and inter-similarity. Through these observations, feature representation and classification can be performed effectively.

\begin{table}[]
	\centering
	\caption{Comparison on the local dataset with noise.}
	\label{tab_metrics_noise}
    \scalebox{0.8}{$
	\begin{tabular}{cccccc}
		\hline
		Methods & \textbf{Accuracy} & \textbf{Precision} & \textbf{Recall} & $\mathbf{F_{1}}$ & \textbf{AUC}   \\ \hline
		\textbf{MLP} & 0.560 & 0.583 & 0.560 & 0.477 & 0.813   \\
		\textbf{SVM} & 0.775 & 0.798 & 0.775 & 0.756 & 0.893   \\
		\textbf{LSTM} & 0.765 & 0.879 & 0.765 & 0.705 & 0.960   \\
		\textbf{Stacked LSTM} & 0.880 & 0.910 & 0.880 & 0.875 & 0.980   \\
		\textbf{DeepBrain} & 0.970 & 0.972 & 0.970 & 0.970 & 0.997   \\ \hline
	\end{tabular}$}
\end{table}
\begin{table}[]
	\centering
	\caption{Comparison on the local dataset without noise.}
	\label{tab_metrics_wo_noise}
    \scalebox{0.8}{$
	\begin{tabular}{cccccc}
		\hline
		Methods & \textbf{Accuracy} & \textbf{Precision} & \textbf{Recall} & $\mathbf{F_{1}}$ & \textbf{AUC}   \\ \hline
		\textbf{MLP} & 0.625 & 0.555 & 0.625 & 0.559 & 0.904   \\
		\textbf{SVM} & 0.790 & 0.783 & 0.790 & 0.785 & 0.886   \\
		\textbf{LSTM} & 0.760 & 0.816 & 0.760 & 0.695 & 0.963   \\
		\textbf{Stacked LSTM} & 0.880 & 0.919 & 0.880 & 0.873 & 0.983   \\
		\textbf{DeepBrain} & 0.975 & 0.977 & 0.975 & 0.975 & 0.998   \\ \hline
	\end{tabular}$}
\end{table}
\begin{figure}[htb]
	\begin{center}
		\begin{tabular}{c}
			\includegraphics[width=0.8\columnwidth]{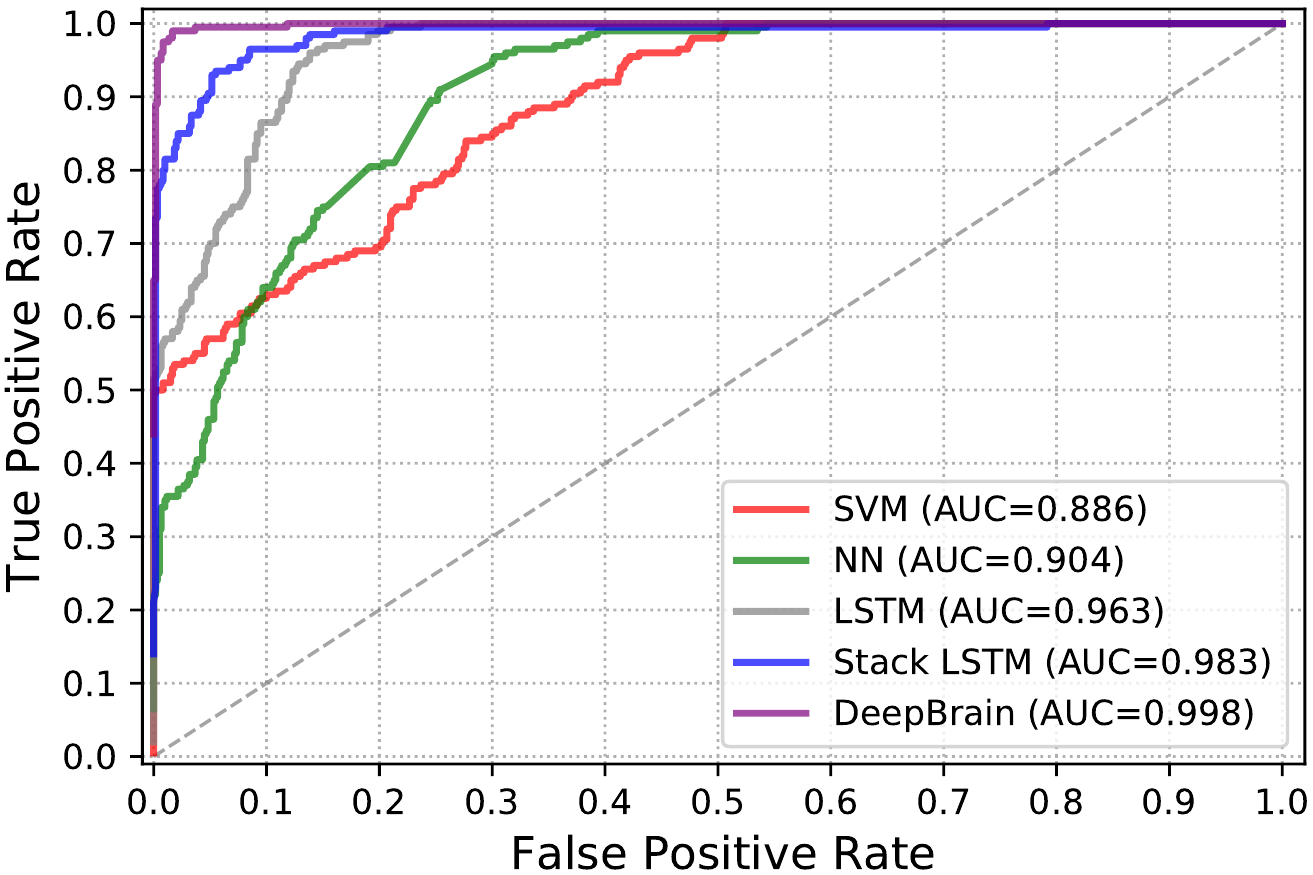} \\
			{\scriptsize (a) ROC with quiet condition.} \\
			\includegraphics[width=0.8\columnwidth]{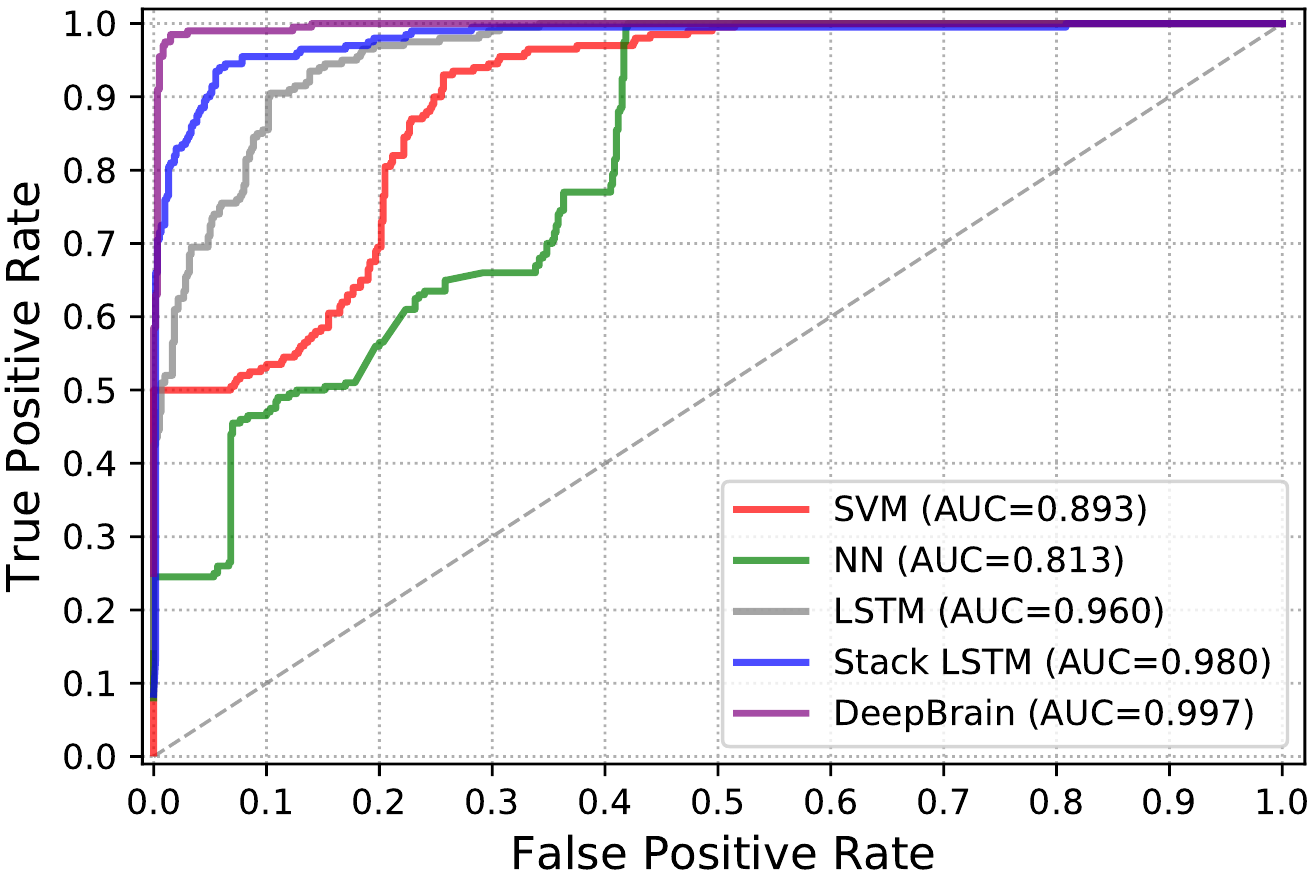}\\
			{\scriptsize (b) ROC with appropriate noise condition.}
		\end{tabular}
	\end{center}
	\caption{The ROC in both noisy and non-noise conditions data as the input of models}
	\label{fig:Figure9}
\end{figure}

\subsection{Overall Comparison with Other Methods}
\begin{figure}[htb]
	\begin{center}
		\begin{tabular}{c}
			\includegraphics[width=0.8\columnwidth]{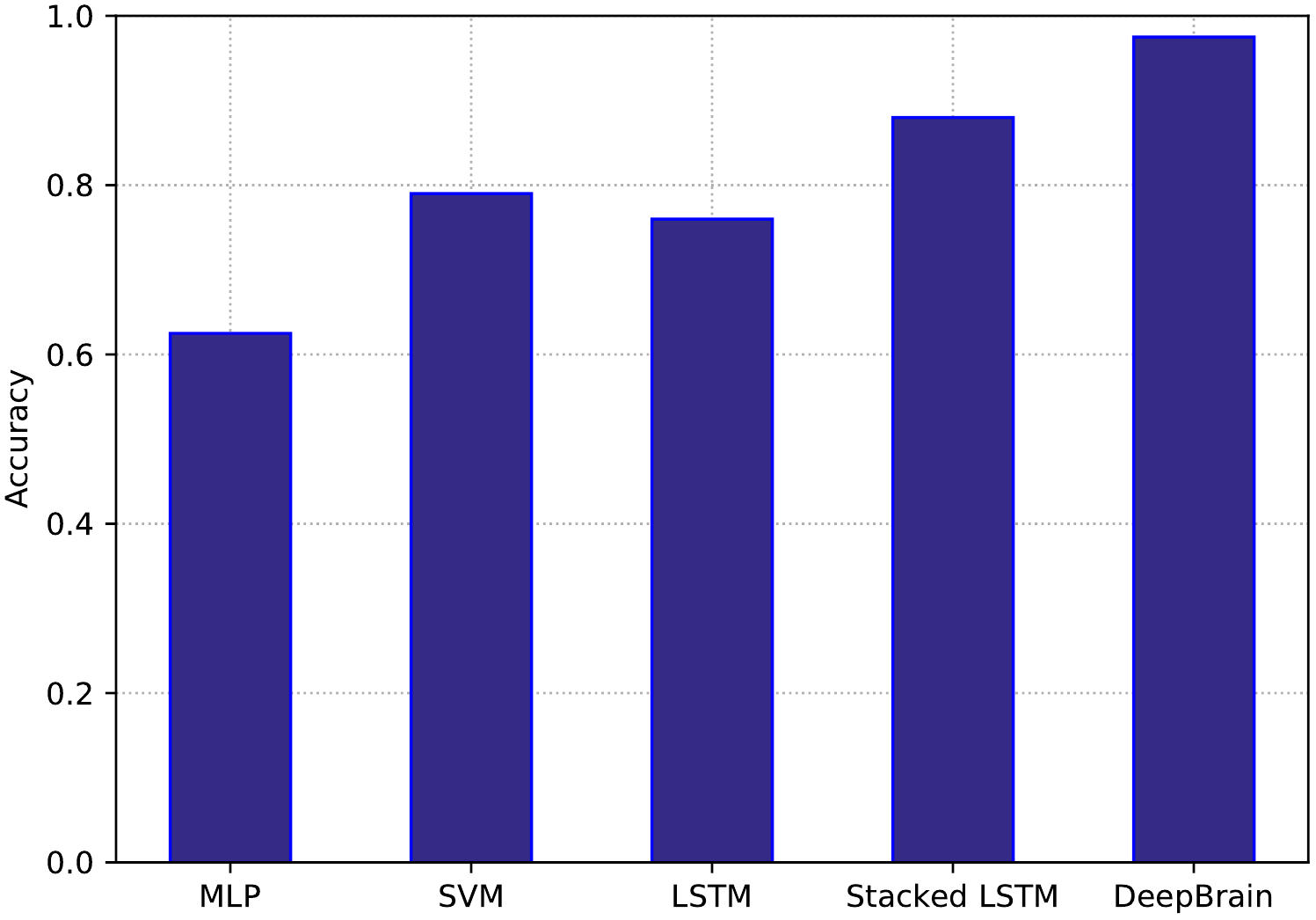} \\
		    {\scriptsize (a) Accuracy comparison.} \\
			\includegraphics[width=0.8\columnwidth]{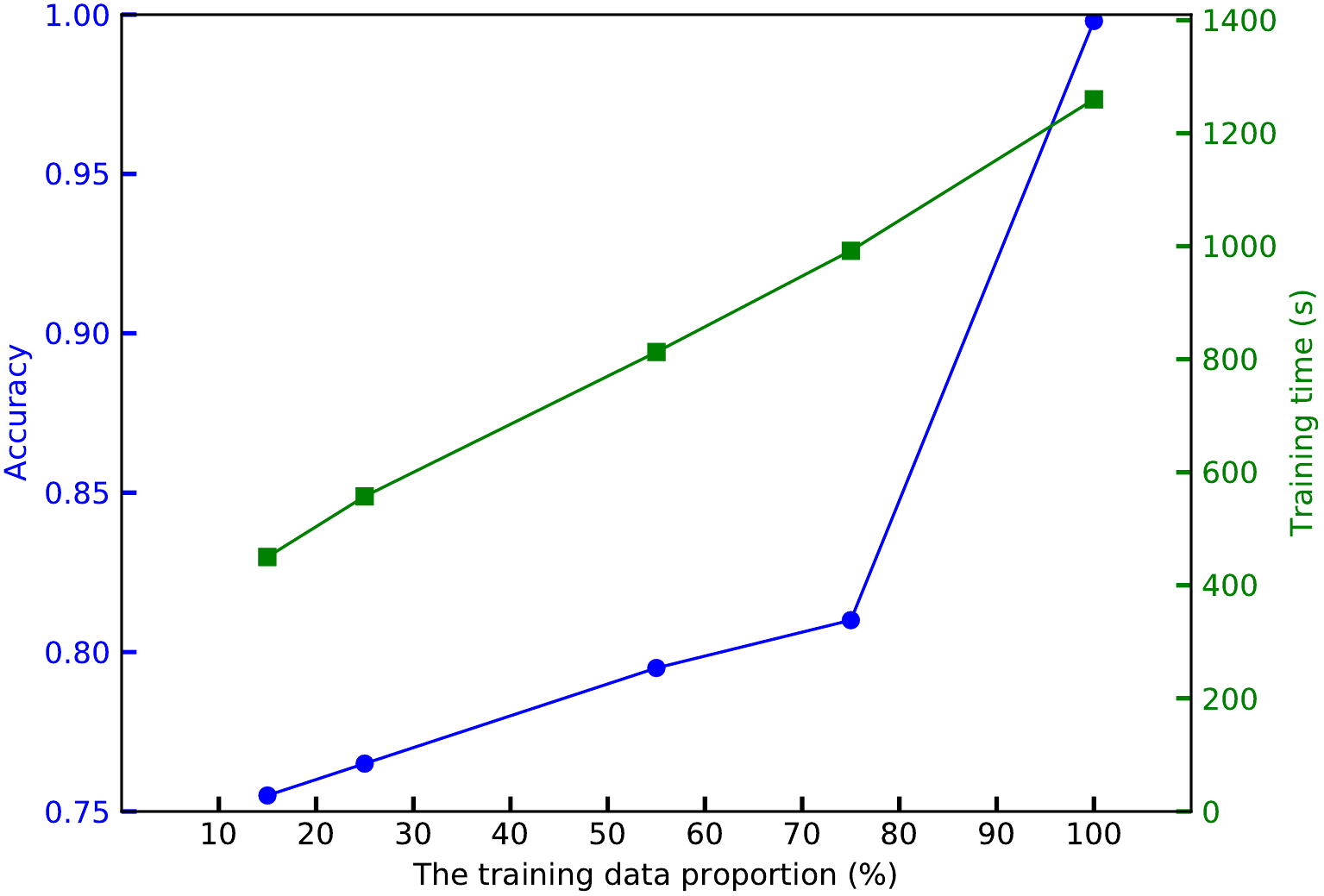}\\
			{\scriptsize (b) Accuracy and training time.}
		\end{tabular}
	\end{center}
	\caption{Comparisons of accuracy and training time.}
	\label{fig:Figure3}
\end{figure}

In this section, we give the performance study and then illustrate the efficiency of our approach by comparing with other methods and other deep learning algorithms. Recall that the designed approach is a hybrid model which uses the LSTM for feature learning and the softmax classifier for intent recognition. In our experiments, the EEG data are randomly divided into two parts: the training dataset and the testing dataset. It should be noted that Brainlink collects EEG data in both noisy and non-noise conditions. We show that our designed model gets the multi-classification accuracy of 0.975 and 0.970 on without noise and noise local dataset, respectively. To take a clear look at the result, we introduce the detailed classification reports in Table~\ref{tab_metrics_wo_noise}. We can observe that the Stacked LSTM with attention enhanced layer is generally better than the general hidden layer in the results of each of metrics. Table~\ref{tab_metrics_noise} shows the metrics on local dataset with noise. It shows that the evaluation metrics on local dataset without noise are better than with noise.
The ROC in both noisy and non-noise conditions data as the input of models is shown in Figure~\ref{fig:Figure9}. The area under DeepBrain curve is larger than the area under the curve of the other three methods, which can also be found from the value of the AUC. It shows that our method is better than the other three methods. According to the definition of ROC curve, we can realize that the ROC unceasingly decreases the threshold of classification, and then count the values of TPR and FPR. Analyzing the ROC curves of DeepBrain in the Figure~\ref{fig:Figure9}, we can find that the TPR value rapidly achieve 0.9 during the process of continuously moving down the threshold value. At the same time, we also analyze the data under appropriate noise in the Figure~\ref{fig:Figure9} (b). For our target group, we believe, the reasonable noise decibels should be below 48 decibels, which is slightly lower than the number of noise decibels when people communicate normally, such as a central air-conditioned room. It means that our approach is much more robust. Although the other three methods perform well, the growth rate of the TPR value is slightly worse than ours. Accuracy comparison between our method and the other three methods are also listed in Figure~\ref{fig:Figure3}.

We compare DeepBrain with SVM, MLP, LSTM, Stacked LSTM.
In addition, the key parameters are listed here: multi-layer perceptron (MLP) (hidden layer node is 30), and LSTM with 32 unit cells. The results illustrate that our designed model achieves the higher accuracy than other methods. Our model also performs better than other deep learning models such as MLP or LSTM. Furthermore, contrasted with the existing EEG classification research which concentrates on binary classification, our designed model runs in multi-class scenario and still achieves a high-level accuracy. To illustrate the advantage of our designed model of robust features from raw EEG data, we also contrast our DeepBrain method with the single deep learning methods MLP and RNN. The experimental results are shown in Figure~\ref{fig:Figure3} (a), where we can notice that our method outperforms MLP, LSTM, SVM and Stacked LSTM in classification accuracy by 35\%, 21.5\%, 18.5\% and 9.5\% respectively. Figure~\ref{fig:Figure3} (b) demonstrates that the accuracy changes along with the training iterations under three categories of feature learning methods, which shows that the designed model converges to its high accuracy in fewer iterations than independent MLP and RNN.

\section{Conclusions}
In this paper, we propose DeepBrain for disabled people's application. We demonstrate a viable technique that the LSTM neural network builds model on normal time series behaviour and then uses prediction to give real-time feedback to our domestic robot. The DeepBrain produces relatively good results on real-world dataset that involves long-term time-dependent and weak time-dependent and is difficult to predict. As compared with MLP, SVM, LSTM, and Stacked LSTM, our model achieves better results, indicating the robustness of our methods.

Future work may consider using different levels of network structure and more accurate EEG collection devices instead of the equipment used in our paper, which can bring more categories of classification and still maintain high level of accuracy since it has more categorizable data and a high-precision network model. In general, the DeepBrain system with its associated methods presents a viable candidate to apply the state-of-the-art AI techniques to the field of HCI applications.

\bibliographystyle{aaai}
\small
\bibliography{sample-bibliography-biblatex}

\begin{thebibliography}{}

\bibitem[\protect\citeauthoryear{Akram, Han, and Kim}{2015}]{Akram2015}
Akram, F.; Han, S.~M.; and Kim, T.-S.
\newblock 2015.
\newblock An efficient word typing p300bci system using a modified t9 interface
  and random forest classifier.
\newblock In {\em Computers in biology and medicine}, volume~56,  30--36.

\bibitem[\protect\citeauthoryear{Calvo and D’Mello}{2010}]{Calvo2010}
Calvo, R.~A., and D’Mello, S.
\newblock 2010.
\newblock Affect detection: An interdisciplinary review of models, methods, and
  their applications,.
\newblock {\em IEEE Trans. Affect. Comput.} 1(1).

\bibitem[\protect\citeauthoryear{Calvo and D’Mello}{2014}]{Jenke2014}
Calvo, R.~A., and D’Mello, S.
\newblock 2014.
\newblock Feature extraction and selection for emotion recognition from eeg.
\newblock {\em IEEE Trans. Affect. Comput.} 5(3).

\bibitem[\protect\citeauthoryear{Graves}{2013}]{Graves2013}
Graves, A.
\newblock 2013.
\newblock Generating sequences with recurrent neural networks.
\newblock {\em arXiv preprint arXiv:1308.0850}.

\bibitem[\protect\citeauthoryear{Gudmundsson \bgroup et al\mbox.\egroup
  }{2007}]{Gudmundsson2007}
Gudmundsson, S.; Runarsson, T.~P.; Sigurdsson, S.; Eiriksdottir, G.; and
  Johnsen, K.
\newblock 2007.
\newblock Reliability of quantitative eeg features.
\newblock In {\em Clin. Neurophysiol.}, volume 118,  2162--2171.

\bibitem[\protect\citeauthoryear{Huang \bgroup et al\mbox.\egroup
  }{2015}]{Huang2015}
Huang, Y.-J.; Wu, C.-Y.; Wong, A. M.-K.; and Lin, B.-S.
\newblock 2015.
\newblock Novel active comb-shaped dry electrode for eeg measurement in hairy
  site.
\newblock {\em IEEE Trans. Biomed. Eng.} 62(1).

\bibitem[\protect\citeauthoryear{Kingma and Ba}{2014}]{Kingma2014}
Kingma, D., and Ba, J.
\newblock 2014.
\newblock Adam: A method for stochastic optimization,.
\newblock {\em arXiv:1412.6980}.

\bibitem[\protect\citeauthoryear{LeCun, Bengio, and Hinton}{2015}]{LeCun2015}
LeCun, Y.; Bengio, Y.; and Hinton, G.
\newblock 2015.
\newblock In {\em Deep learning}, volume 521,  436--444.
\newblock Nature.

\bibitem[\protect\citeauthoryear{Mauss and Robinson}{2009}]{Mauss2009}
Mauss, I.~B., and Robinson, M.~D.
\newblock 2009.
\newblock Measures of emotion: A review.
\newblock In {\em Cogn. Emotion}, volume~23,  209--237.

\bibitem[\protect\citeauthoryear{Muller}{2003}]{Muller2003}
Muller, M.~J.
\newblock 2003.
\newblock Participatory design: the third space in hci.
\newblock In {\em Human-computer interaction: Development process},  165--185.

\bibitem[\protect\citeauthoryear{Nguyen \bgroup et al\mbox.\egroup
  }{2015}]{Nguyen2015}
Nguyen, T.; Nahavandi, S.; Khosravi, A.; Creighton, D.; and Hettiarachchi, I.
\newblock 2015.
\newblock Eeg signal analysis for bci application using fuzzy system.
\newblock {\em 2015 International Joint Conference on Neural Networks (IJCNN)}.

\bibitem[\protect\citeauthoryear{O.R.Pinheiro \bgroup et al\mbox.\egroup
  }{2016}]{Pinheiro2016}
O.R.Pinheiro; J.R.deSouza; L.R.Alves; and M.Romero.
\newblock 2016.
\newblock Wheelchair simulator game for training people with severe
  disabilities.
\newblock {\em IEEE}.

\bibitem[\protect\citeauthoryear{{Qin} \bgroup et al\mbox.\egroup
  }{2018}]{Qin2018}
{Qin}, Z.; {Wu}, D.; {Xiao}, Z.; {Fu}, B.; and {Qin}, Z.
\newblock 2018.
\newblock Modeling and analysis of data aggregation from convergecast in mobile
  sensor networks for industrial iot.
\newblock {\em IEEE Transactions on Industrial Informatics} 14(10):4457--4467.

\bibitem[\protect\citeauthoryear{Schmidhuber and
  Hochreiter}{1997}]{Schmidhuber1997}
Schmidhuber, J., and Hochreiter, S.
\newblock 1997.
\newblock Long short-term memory.
\newblock In {\em Neural computation}, volume~9,  1735--1780.

\bibitem[\protect\citeauthoryear{Schmidhuber, Graves, and
  Fernandez}{2007}]{Schmidhuber2007}
Schmidhuber, J.; Graves, A.; and Fernandez, S.
\newblock 2007.
\newblock Multi-dimensional recurrent neural networks.
\newblock In {\em International Conference on Artificial Neural Networks},
  549--558.

\bibitem[\protect\citeauthoryear{Stober \bgroup et al\mbox.\egroup
  }{2015}]{Stober2015}
Stober, S.; Sternin, A.; Owen, A.~M.; and Grahn, J.~A.
\newblock 2015.
\newblock Deep feature learning for eeg recordings.
\newblock {\em arXiv preprint arXiv:1511.04306}.

\bibitem[\protect\citeauthoryear{Williams \bgroup et al\mbox.\egroup
  }{2015}]{Williams2015}
Williams, M.~A.; Roseway, A.; O’Dowd, C.; Czerwinski, M.; and Morris, M.~R.
\newblock 2015.
\newblock Swarm: An actuated wearable for mediating affect.
\newblock {\em Proc. 9th ACM Int. Conf. Tangible Embedded Embodied
  Interaction}.

\bibitem[\protect\citeauthoryear{{Wu} \bgroup et al\mbox.\egroup
  }{2020}]{Wu2020}
{Wu}, D.; {Nie}, X.; {Asmare}, E.; {Arkhipov}, D.~I.; {Qin}, Z.; {Li}, R.;
  {McCann}, J.~A.; and {Li}, K.
\newblock 2020.
\newblock Towards distributed sdn: Mobility management and flow scheduling in
  software defined urban iot.
\newblock {\em IEEE Transactions on Parallel and Distributed Systems}
  31(6):1400--1418.

\bibitem[\protect\citeauthoryear{{Zhang} \bgroup et al\mbox.\egroup
  }{2010}]{Zhang2010}
{Zhang}, Y.; {Bao}, L.; {Yang}, S.; {Welling}, M.; and {Wu}, D.
\newblock 2010.
\newblock Localization algorithms for wireless sensor retrieval.
\newblock {\em The Computer Journal} 53(10):1594--1605.

\bibitem[\protect\citeauthoryear{Zhang \bgroup et al\mbox.\egroup
  }{2017}]{Zhang2017}
Zhang, X.; Yao, L.; Sheng, Q.~Z.; Kanhere, S.~S.; Gu, T.; and Zhang, D.
\newblock 2017.
\newblock Converting your thoughts to texts: Enabling brain typing via deep
  feature learning of eeg signals.
\newblock {\em arXiv:1709.08820}.

\bibitem[\protect\citeauthoryear{Zhang \bgroup et al\mbox.\egroup
  }{2018}]{Lina2018}
Zhang, X.; Yao, L.; Kanhere, S.; Liu, Y.; Gu, T.; and Chen., K.
\newblock 2018.
\newblock Mindid: Person identification from brain waves through
  attention-based recurrent neural network.
\newblock {\em ACM International Joint Conference on Pervasive and Ubiquitous
  Computing (Ubicomp 2018)}.

\bibitem[\protect\citeauthoryear{Zheng \bgroup et al\mbox.\egroup
  }{2018}]{WeiLongZheng2018}
Zheng, W.-L.; Liu, W.; Lu, Y.; Lu, B.-L.; and Cichocki, A.
\newblock 2018.
\newblock Emotionmeter: A multimodal framework for recognizing human emotions.
\newblock {\em IEEE Transactions on Cybernetics}.

\end{thebibliography}

\end{document}